\documentclass[]{aa}
\usepackage{todonotes}
\usepackage{graphicx}
\usepackage{natbib}
\usepackage[varg]{txfonts}
\usepackage{amsmath}
\usepackage{multirow}
\usepackage{color}
\usepackage{float}
\usepackage{lscape}
\usepackage{xcolor}
\usepackage{url}
\usepackage[switch]{lineno}
\usepackage{hyperref} 

\newcommand\kms{\ensuremath{\mbox{km}\,\mbox{s}^{-1}}}
\newcommand\Teff{\ensuremath{T_\mathrm{eff}}~}
\newcommand\logg{\ensuremath{\log g}~}

\newcommand\met{\ensuremath{\mbox{[Fe/H]}}}
\newcommand\fei{\ion{Fe}{i}}

\newcommand\sii{\ion{Si}{i}}

\newcommand\nai{\ion{Na}{i}}
\newcommand\naii{\ion{Na}{ii}}

\newcommand\mgi{\ion{Mg}{i}}
\newcommand\mgii{\ion{Mg}{ii}}

\newcommand\ali{\ion{Al}{i}}
\newcommand\alii{\ion{Al}{ii}}

\bibpunct{(}{)}{;}{a}{}{,}

\csname @openrightfalse\endcsname
\hypersetup{draft}
\begin{document}

\title{An empirical recipe for inelastic hydrogen-atom collisions in non-LTE calculations\thanks{The tables in the Appendix A.1. to A.4. are only available in electronic form at the CDS via anonymous ftp to cdsarc.u-strasbg.fr (130.79.128.5) or via http://cdsweb.u-strasbg.fr/cgi-bin/qcat?J/A+A/}}
\author{R. Ezzeddine\inst{1}$^{,}$\inst{2}$^{,}$\inst{3}\thanks{email: \texttt{ranae@mit.edu}} \and
T. Merle\inst{4} \and
B. Plez\inst{1} \and
M. Gebran\inst{5}\and
F. Th\'{e}venin\inst{6} \and M. Van der Swaelmen\inst{4}}

\institute{
Laboratoire Univers et Particules de Montpellier, Universit\'e
de Montpellier, CNRS, UMR 5299, Montpellier, France
\and
Joint Institute for Nuclear Astrophysics, Center for the Evolution of the Elements, East Lansing, MI 48824, USA
\and
Kavli Institute for Astrophysics and Space Research, Massachusetts Institute of Technology, 
Cambridge, MA 02139, USA
\and
Institut d'Astronomie et d'Astrophysique, Universit\'{e} Libre de Bruxelles, CP 226, Boulevard du Triomphe, 1050 Brussels, Belgium
\and
Department of Physics and Astronomy, Notre Dame University-Louaize, PO Box 72, Zouk Mika\"el, Lebanon
\and 
Universit\'{e} C\^{o}te d'Azur, Observatoire de la C\^{o}te d'Azur, CNRS, UMR7293, CS 34229, F-06304 Nice Cedex 4}
\date{Received / Accepted}

\titlerunning{An empirical recipe for inelastic hydrogen-atom collisions in non-LTE calculations}
\authorrunning{Ezzeddine et al.}

\abstract{Determination of high-precision abundances of late-type stars has been and always will be an important goal of spectroscopic studies, which requires accurate modeling of their stellar spectra with non-local thermodynamic equilibrium (NLTE) radiative transfer methods. This entails using up-to-date atomic data of the elements under study, which are still subject to large uncertainties.}
{We investigate the role of hydrogen collisions in NLTE spectral line synthesis, and introduce a new general empirical recipe to determine inelastic charge transfer (CT) and bound-bound hydrogen collisional rates.
 This recipe is based on fitting the energy functional dependence of published quantum collisional rate coefficients of several neutral elements (\ion{Be}{I}, \ion{Na}{I}, \ion{Mg}{I}, \ion{Al}{I}, \ion{Si}{I} and \ion{Ca}{I}) using simple polynomial equations.}
{We perform thorough NLTE abundance calculation tests using our method for four different atoms, Na, Mg, Al and Si, for a broad range of stellar parameters. We then compare the results to calculations computed using the published quantum rates for all the corresponding elements. We also compare to results computed using excitation collisional rates via the commonly used Drawin equation for different fudge factors, S$_{\mathrm{H}}$, applied.}
{We demonstrate that our proposed method is able to reproduce the NLTE abundance corrections performed with the quantum rates for different spectral types and metallicities for representative \nai\ and \ali\ lines to within $\leq0.05$\,dex and $\leq0.03$\,dex, respectively. For \mgi\ and \sii\ lines, the method performs better for the cool giants and dwarfs, while larger discrepancies up to 0.2\,dex could be obtained for some lines for the subgiants and warm dwarfs. We obtained larger NLTE correction differences between models incorporating Drawin rates relative to the quantum models by up to 0.4\,dex. These large discrepancies are potentially due to ignoring either or both CT and ionization collisional processes by hydrogen in our Drawin models.}
{Our general empirical fitting method (EFM) for estimating hydrogen collision rates performs well in its ability to reproduce, within narrow uncertainties, the abundance corrections computed with models incorporating quantum collisional rates. It performs generally best for the cool and warm dwarfs, with slightly larger discrepancies obtained for the giants and subgiants.
It could possibly be extended in the future to transitions of the same elements for which quantum calculations do not exist, or, in the absence of published quantum calculations, to other elements as well.}
 \maketitle 

\section{Introduction}
\label{intro}
In the era of large-scale surveys tracing the history of stellar populations (e.g., RAVE, \citealp{steinmetz2006}; APOGEE, \citealp{allendep2008}; \textit{Gaia}-ESO, \citealp{gilmore2012}, etc.), there is an important need for high-accuracy, detailed chemical composition determinations.
Most stellar spectroscopic analyses adopt the assumption of local thermodynamic equilibrium (LTE).
This assumption is however not always valid, especially when collisions are not frequent enough to ensure excitation and ionization equilibrium of the atomic populations. Collisions in cool stellar atmospheres are mainly due to electrons and hydrogen atoms. Although electrons have larger thermal velocities than hydrogen atoms by a factor of approximately 43, they can be less numerous than hydrogen atoms in late-type stellar atmospheres \citep{lambert1993}. In fact, in low-metallicity stars, electron densities are much lower (by the order of $10^{-4}$) and non-LTE (NLTE) effects are expected to grow stronger, as well as in giants' and supergiants' rarefied atmospheres \citep{thevenin1999, asplund2005, lind2011}. 

NLTE modeling requires a great deal of atomic data for each element under consideration. Large experimental and theoretical efforts have been devoted to measuring and calculating energy levels, $gf$-values, and broadening parameters (e.g., \citealp{anstee1995,castelli2010,petkur2015,abo2015}), for example. Hydrogen collision rates remain, however, a major source of uncertainty in NLTE calculations for cool stars \citep{asplund2005,barklem2010}. Quantum mechanical collision rates have been computed for a small number of atoms over the past decade including Be \citep{yakovleva2016}, Na \citep{barklem2010}, Mg \citep{belyaev2012,guitou2015}, Al \citep{belyaev2013}, Si \citep{belyaev2014}, K \citep{yakovleva2018}, Ca \citep{barklem2016,mitrushchenkov2017,belyaev2017b}, Mn \citep{belyaev2017Mn} and Rb \citep{yakovleva2018}.

In the absence of such calculations for other atoms,
the Drawin approximation \citep{drawin1968,drawin1969a,drawin1969b,steenbock1984,lambert1993} is customarily used to estimate the collisional cross-sections.
This approximation was originally derived from the classical \citet{thomson1912} $e^{-}$ + atom ionization rate equation.
The Drawin equation has been applied to allowed bound-bound ($b-b$) and ionization bound-free ($b-f$) transitions, where it has been  shown to overestimate the collisional rates by orders of magnitude \citep{barklem2010}.
A fudge-factor, S$_{\mathrm{H}}$, is usually applied in most studies to try to compensate for this overestimation. Several S$_{\mathrm{H}}$-factors have been proposed for different atoms. As an example, for \ali, values between 0.002 and 0.3 have been suggested (e.g., S$_{\mathrm{H}}=0.1$ \citealp{mashonkina2016}, S$_{\mathrm{H}}=0.002$ \citealp{gehren2004,Baumueller1996,Baumueller1997}, S$_{\mathrm{H}}=0.1$ \citealp{andrievsky2008} and S$_{\mathrm{H}}=0.3$ \citealp{steenbock1992}).

Recently, charge transfer (CT) rates, corresponding to the ion-pair production and mutual-neutralization
processes: $A$ + H $\rightleftharpoons$ $A^{+}$ + H$^{-}$ (A denoting an atom other than H), have been shown to be larger than the $b-b$ rates \citep{barklem2010}. 
During a collision, the valence electron of the atom $A$ has the potential to tunnel to the H atom at an avoided crossing. Later,  at a different avoided crossing, it may tunnel back to another 
covalent molecular level, leading to a different final state of the atom $A$, and an excitation or de-excitation of the atom. The electron 
may also stay with the H atom, leading to an ion-pair production. The former process is more likely to have a small-valued cross-section as it implies passing two avoided crossings, the higher-lying  being passed with a small transition probability \citep{barklem2011}. The CT rates therefore play an important, and in some cases, even dominating role in the collisional processes \citep{lind2011,osorio2015,guitou2015} and should therefore be included in NLTE calculations. In \citet{ezzeddine2016}, we used the Drawin formula to estimate CT rates, 
in the absence of any other approximation. We defined  two scaling-factors S$_{\mathrm{H}}$ for $b-b$ and $b-f$, and S$_{\mathrm{H}}$(CT) for
CT rates for \fei\ transitions, and tried to calibrate them using benchmark star spectra ($\alpha$ Cen A, HD\,140283 and the Sun). Our method used a $\chi^2$-minimization of differences between observed and calculated iron-line equivalent widths (EWs) for different sets of S$_{\mathrm{H}}$ and S$_{\mathrm{H}}$(CT).
We could not find a combination of S$_{\mathrm{H}}$ and S$_{\mathrm{H}}$(CT) that would work for the three stars.
This difficulty of the Drawin formula to reproduce the behavior and magnitude of hydrogen collision rates, independent of the stellar and atomic models used, calls for another approach, as long as no quantum calculation are available.

This is the motivation for our work, which introduces an empirical fitting method (EFM) to estimate the hydrogen collision rates for both $b-b$ and CT processes. It is based on the observation that the existing well determined quantum rates all behave similarly, and can be described with simple fitting functions.
Recent studies by \citet{belyaev2017neutral} and \citet{belyaev2017ionic} also used such observations to derive general recipes for estimating neutral and ionic hydrogen collisional rates for $b-b$ and CT rate coefficients based on a simplified approach of the Landau-Zener R-matrix model. Their results were published during the course and after the submission of this work. In their recipes they estimate $b-b$ and CT reduced rate coefficients of dominant transitions as a function of collision energies and statistical probabilities of the relevant atomic levels, which they tested on K for neutral rates and Ba for ionic rates. We performed comprehensive tests of our method on four atoms belonging to different groups in the periodic table for which quantum rates have been calculated, namely Na, Mg, Al and Si. We built model atoms for each of these elements and  implemented different recipes for H collisions, including our empirical fitting recipe, published quantum rates as well as rates computed using the Drawin equation. We then use them to perform and compare NLTE calculations for a large range of stellar parameters.

This article is divided as follows. In Section~\ref{efm}, we introduce the empirical fitting method and its application to estimate the hydrogen collision rates for both CT and de-excitation rates. In Section~\ref{atoms_tests}, we put our method to the test on four elements for which we built new model atoms. In Section~\ref{Sec:results}, we compare the line profiles and NLTE abundance corrections obtained by the different models and present our results. Finally, conclusions are presented in Section~\ref{conclusion}.

\section{Introducing the general empirical recipe}\label{efm}
\begin{table*}
\begin{center}
 \begin{tabular}{c | c c c c c c  ||  r  r  r  r  r  r  r}
 \hline
 \hline
 & \multicolumn{6}{c||}{CT} & \multicolumn{7}{c}{$b-b$}\\
 \hline
  El &  $a_{11}$ & $a_{10}$ & $a_{2}$ & $\frac{1}{\Delta E_{0}}$ & $\alpha$ & $\log Q_{0}$ & $b_{31}$ & $b_{30}$ & $b_{21}$ & $b_{20}$ & $b_{11}$ & $b_{10}$ & $b_{0}$\\
 \hline
\ion{Be}{i}   & $-1.52$ & $5.12$ & $-2.81$ & $0.71$ & $-0.48$ & $-7.39$ & $-0.13$ & $0.09$  & 0.86   & $-0.73$  & $-1.18$ & $0.66$   & $-10.50$\\
\ion{Na}{i}   & $-1.60$ & $7.99$ & $-1.97$ & $0.67$ & $-0.57$ & $-7.50$ & $-0.03$ & $-0.24$ & $0.48$ & $1.84$  & $-1.69$ & $-5.75$   &  $-9.84$  \\
\ion{Mg}{i}   & $-0.16$ & $3.82$ & $-1.10$ & $0.69$ & $-0.25$ & $-7.40$ & $0.09$ & $0.12$ & $-0.45$ & $-0.85$   & $0.31$ & $0.77$   &  $-10.17$  \\
\ion{Al}{i}   & $-0.60$ & $2.88$ & $-2.70$ & $0.83$ & $-0.18$ & $-7.36$ & $-0.88$ & $0.15$  & $5.89$ & $-1.48$ & $-8.43$ & $0.23$  & $-10.87$  \\
\ion{Si}{i}   & $-1.20$ & $4.90$ & $-3.52$ & $0.75$ & $-0.34$ & $-7.47$ & $-0.08$ & $-0.09$  & $0.70$ & $0.75$  & $-1.36$ & $-1.93$   &  $-9.76$ \\
\ion{Ca}{i}   &$-1.65$  & 9.44   & $-2.29$ & $0.71$ & $-0.32$ & $-6.98$ & $0.19$ & 0.40    & -0.63   & $-2.80$  & $0.54$ &  3.57    & $-11.30$\\
 \hline 
 \end{tabular}
\end{center}
\caption{CT and $b-b$ hydrogen collision rate fitting coefficients obtained for the neutral species of Be, Na, Mg, Al, Si and Ca.}
\label{fit_param_all}
\end{table*}

\begin{figure*}
\begin{center}
\hspace*{-1cm}
 \includegraphics[scale=0.35]{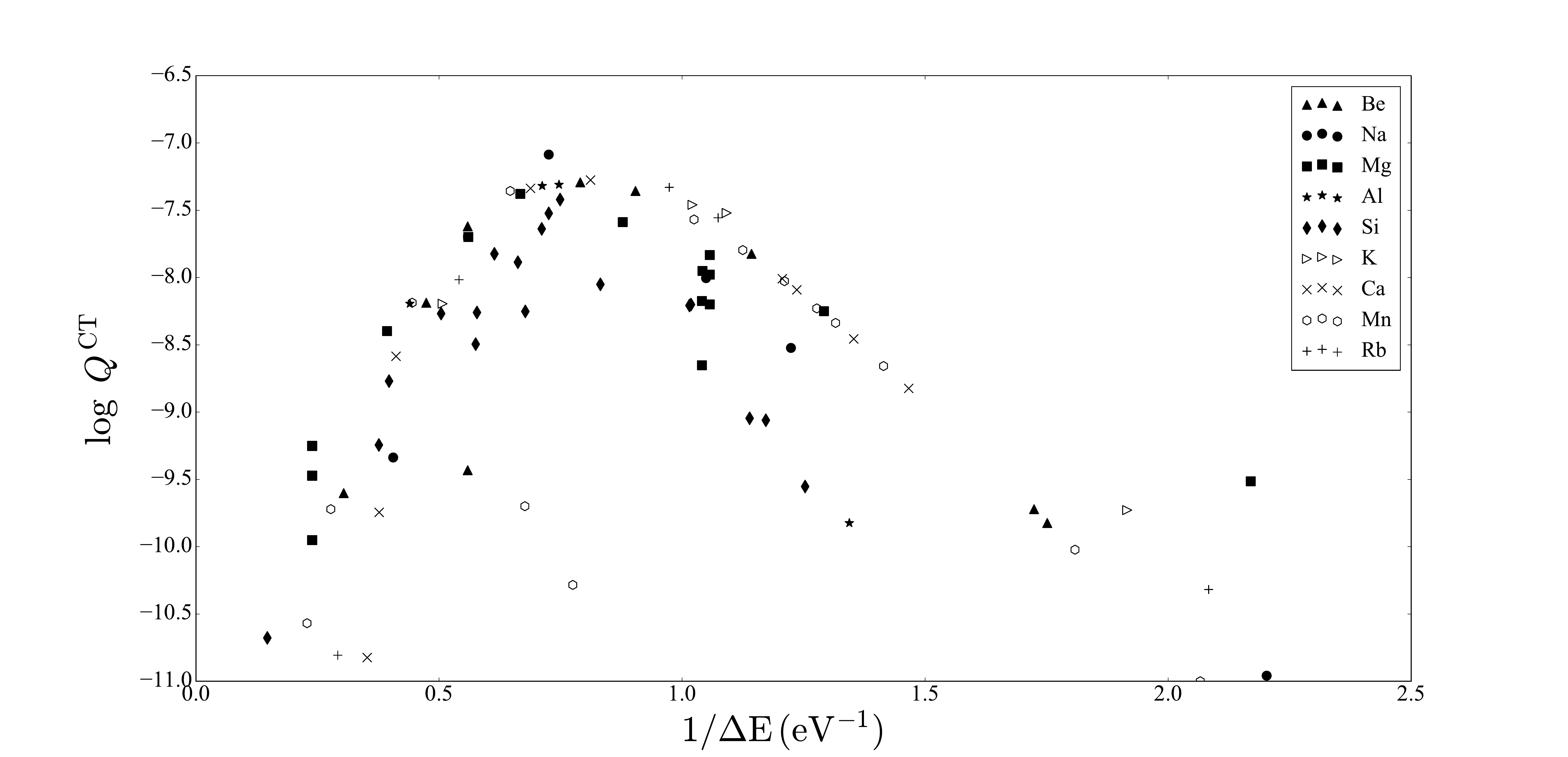}
 \end{center}
 \caption{CT rate coefficients (in logarithmic scale), $\log Q^{\mathrm{CT}}$, as a function of inverse 
 transition energies $\frac{1}{\Delta E}$ at $T = 6000$\,K, for \ion{Be}{i}, \ion{Na}{i}, \ion{Mg}{i}, \ion{Al}{i}, \ion{Si}{i}, \ion{K}{i}, \ion{Ca}{i}, \ion{Mn}{i} and \ion{Rb}{i} from the quantum calculations of \citet{yakovleva2016},
 \citet{barklem2010}, \citet{belyaev2012} and \citet{guitou2015}, \citet{belyaev2013}, \citet{belyaev2014}, \citet{yakovleva2018}, \citet{belyaev2017b}, \citet{belyaev2017} and \citet{yakovleva2018},
 respectively.}
 \label{plot_qm_ce_all}
\end{figure*}
Our empirical recipe to estimate the hydrogen collisional rates for both CT and (de-)excitation processes is motivated by the similar behavior of the quantum-mechanical hydrogen collision rates for several neutral species of atoms including Be \citep{yakovleva2016}, Na \citep{barklem2010}, Mg \citep{belyaev2012,guitou2015}, Al \citep{belyaev2013}, Si \citep{belyaev2014}, K \citep{yakovleva2018}, Ca \citep{barklem2016,mitrushchenkov2017,belyaev2017b}, Mn \citep{belyaev2017Mn} and Rb \citep{yakovleva2018}. This similarity is shown in Figure~\ref{plot_qm_ce_all}, where the CT rates at $T=6000$\,K are shown for these nine atoms.

We chose to plot the  logarithm of the Maxwellian averaged cross-sections $Q = Q^{\mathrm{CT}}$ (cgs units) as a function of inverse transition energy $1/\Delta E$ (eV$^{-1}$), where $\Delta E = E_{up} - E_{low}$
is the energy difference between the upper and the lower level of the transition. As the upper state consists of H$^-$ + $A^+$, its energy is the iron ionization potential decreased by 0.754\,eV, the H$^-$ electron affinity. Figure~\ref{plot_qm_ce_all} shows that the rates of all nine elements have a peak value of 
$Q^{\mathrm{CT}} \sim10^{-7}$\,cm$^3$\,s$^{-1}$ arising at $1/\Delta E$ between $0.5$ and $1$\,eV$^{-1}$. The empirical fitting method (EFM) is based on fitting these rates as a function of $\Delta E$ using simple general recipes, as explained in detail below. The hydrogen rate coefficients for K, Mn and Rb were published during the course of this work, and were thus excluded from the fitting process and analysis hereafter. However, their rates as shown in Figure~\ref{plot_qm_ce_all}, behave similarly to the other six atoms. Their exclusion from the analysis should therefore not affect the final results of this work.

\subsection{Charge transfer rates}\label{ct}
\begin{figure*}
\begin{center}
\hspace*{-1cm}
 \includegraphics[scale=0.35]{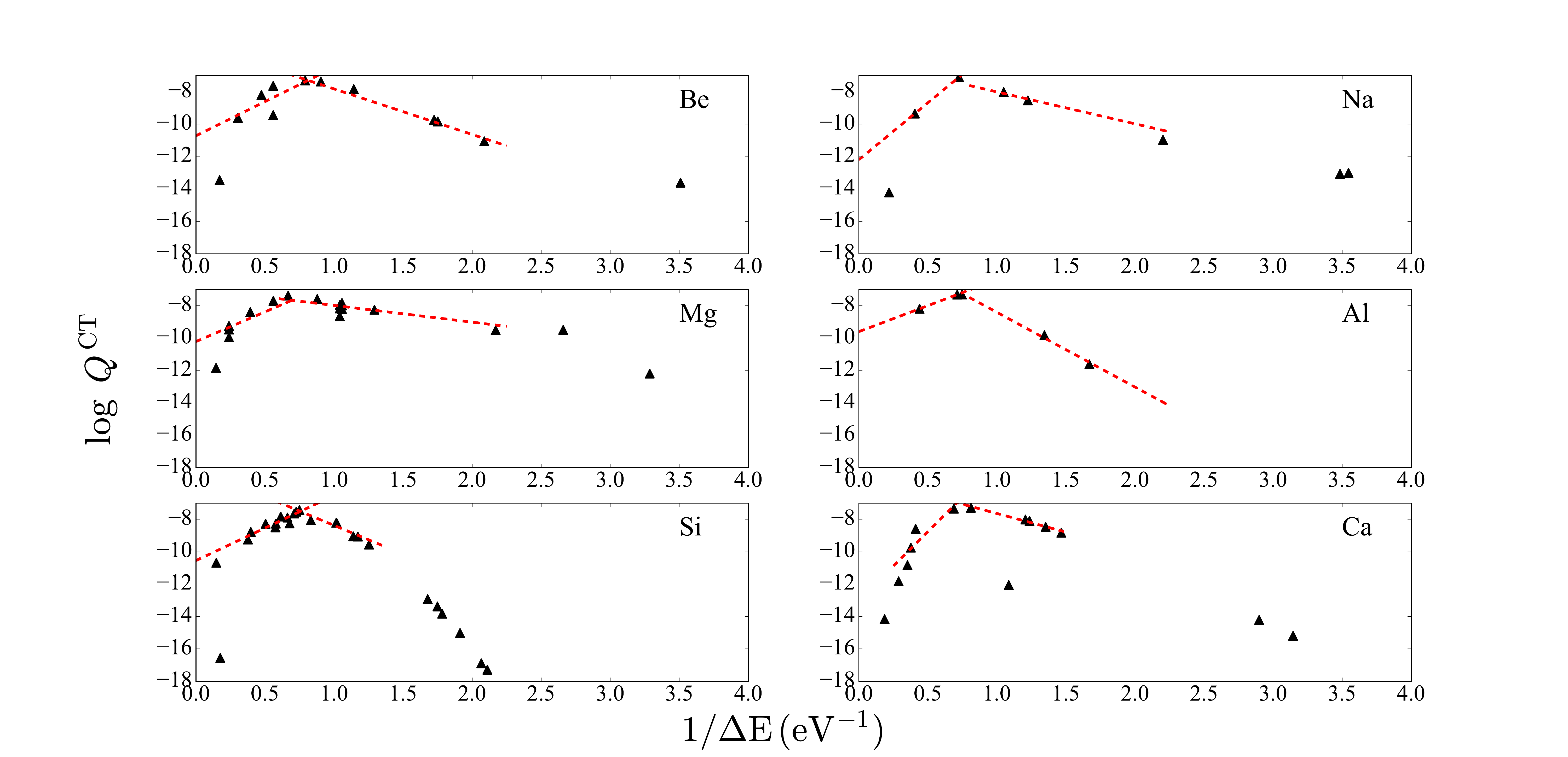}
 \end{center}
 \caption{CT rate coefficients $\log Q^{\mathrm{CT}}$ at $T = 6000$\,K, for \ion{Be}{i}, \ion{Na}{i}, \ion{Mg}{i}, \ion{Al}{i}, \ion{Si}{i} and \ion{Ca}{i} (filled triangles) from the quantum calculations of \citet{yakovleva2016},
 \citet{barklem2010}, \citet{belyaev2012} and \citet{guitou2015}, \citet{belyaev2013}, \citet{belyaev2014} and \citet{belyaev2017b}, respectively. The polynomial fits as per the EFM (dashed red lines) are also displayed on the plots.}
 \label{plot_qm_ce_fits_all}
\end{figure*}
 Figure~\ref{plot_qm_ce_fits_all} shows the fits to the logarithm of the mutual-neutralization (downward CT) rate coefficients, $\log Q^{\mathrm{CT}}$, of \ion{Be}{i}, \ion{Na}{i}, \ion{Mg}{i}, \ion{Al}{i}, \ion{Si}{i} and \ion{Ca}{i} from 
 \citet{yakovleva2016}, \citet{barklem2010}, \citet{belyaev2012} and \citet{guitou2015}, \citet{belyaev2013}, 
 \citet{belyaev2014} and \citet{belyaev2017b}, respectively. 
 We only include the rates at $T=6000$\,K, as they show a similar behavior at other temperatures. The temperature dependence is discussed further below.

We chose an inverse energy scale representation to allow simple linear fits of most of the data. Rates lying more than four orders of magnitude below the peak were ignored, as such rates will not contribute significantly to the NLTE calculations \citep{barklem2011}.
We are thus able to describe the CT rate coefficients, $\log Q^{\mathrm{CT}}$, of all six elements as a function of
$1/\Delta E$ with the following recipe: 

\begin{equation} \label{gen_fit_eqn_ce} 
 \log \, Q^{\mathrm{CT}} = 
 \begin{cases}
   a_{1} \bigg ( \frac{1}{\Delta E} - \frac{1}{\Delta E_{0}} \bigg) + \log \, Q_{\mathrm{max}} \, \, \, \, \, \, \, \, \text{\textit{for}   } \, \, \, \frac{1}{\Delta E} < \frac{1}{\Delta E_{0}} \\
   a_{2} \bigg ( \frac{1}{\Delta E} - \frac{1}{\Delta E_{0}} \bigg) + \log \, Q_{\mathrm{max}} \, \, \, \, \, \, \, \, \text{\textit{for}   } \, \, \, \frac{1}{\Delta E} > \frac{1}{\Delta E_{0}} \\
 \end{cases}
\end{equation}

where $a_{1}$ and $a_{2}$ are  the positive and negative slopes of the linear fits, respectively,
and $Q_{\mathrm{max}}$ is the maximum peak rate at $\frac{1}{\Delta E_{0}}$. The temperature dependence of the CT fitting parameters $a_{1}$, $a_{2}$, $\frac{1}{\Delta E}$ and $\log Q_{\mathrm{max}}$
were also investigated. As an illustration, the temperature dependence is shown  for the case of Si in the upper panels of Figure~\ref{param_temp_si}. The coefficient  $a_1$ is  roughly linearly temperature-dependent, while $\log Q_{\mathrm{max}}$ depends
linearly on $\log T$. The coefficients $a_{2}$ and $\frac{1}{\Delta E_{0}}$ do not vary significantly with temperature. It follows that $a_{1}$ and $\log Q_{\mathrm{max}}$ can be written as a function of temperature $T$ as:

\begin{equation}\label{temp_dep_ce_eqn}
 \begin{cases}
 a_{1} = a_{11} \, \bigg(\frac{T}{10^{4}}\bigg) + a_{10}\\
 \log Q_{\mathrm{max}} = \alpha \log (\frac{T}{10^4}) + \log Q_{0}
 \end{cases}
.\end{equation}

Coefficients $a_{11}$, $a_{10}$, $\alpha$ and $\log Q_{0}$, as well as $a_{2}$ and $1/\Delta E_{0}$ obtained for the six elements
are shown in Table~\ref{fit_param_all}.
The average position in energy of the maximum rate, 
$\log Q_0 = -7.35 \pm 0.37$, 
is well defined at  $1/\Delta E_0 = 0.73 \pm 0.10\,{\rm eV}^{-1}$, with a temperature exponent $\alpha = -0.35 \pm 0.22$.
More scatter is found for the $a$ coefficients, with an average $a_{10} = 5.7\pm 3.7$, $a_{11} = -1.1\pm 0.9$, and 
$a_{2} = -2.4\pm 1.2$. 
The temperature dependence is weak for all six elements, with a variation of $\log Q^{\mathrm{CT}}$ by a factor of about $1.5$ in the temperature range $3000 - 10000\,{\rm K}$. The slope at small energy difference, $a_2$, is independent of the temperature, and at larger energy difference $a_1$ decreases by less than $15\%$.
The effects of the temperature dependence of the rates on NLTE abundance calculations are further discussed in Section~\ref{sec:temp_dep}.

For the six elements, the functional dependencies on energy and temperature of the rates, and hence of the fitting coefficients, are very similar; the peak value of the CT rates, their position, and even the slopes with energy are quite close.
The reason for this similarity in the  behavior of the rates is the presence of avoided crossings of the AH molecular potentials 
and the A$^+$H$^-$ ionic potential.
The peaks shown for all the atoms at approximately the same transition energies correspond to transitions occurring at optimal internuclear distances \citep{barklem2011,barklem2018}.

\begin{figure*}
\begin{center}
\hspace*{-1cm}
 \includegraphics[scale=0.35]{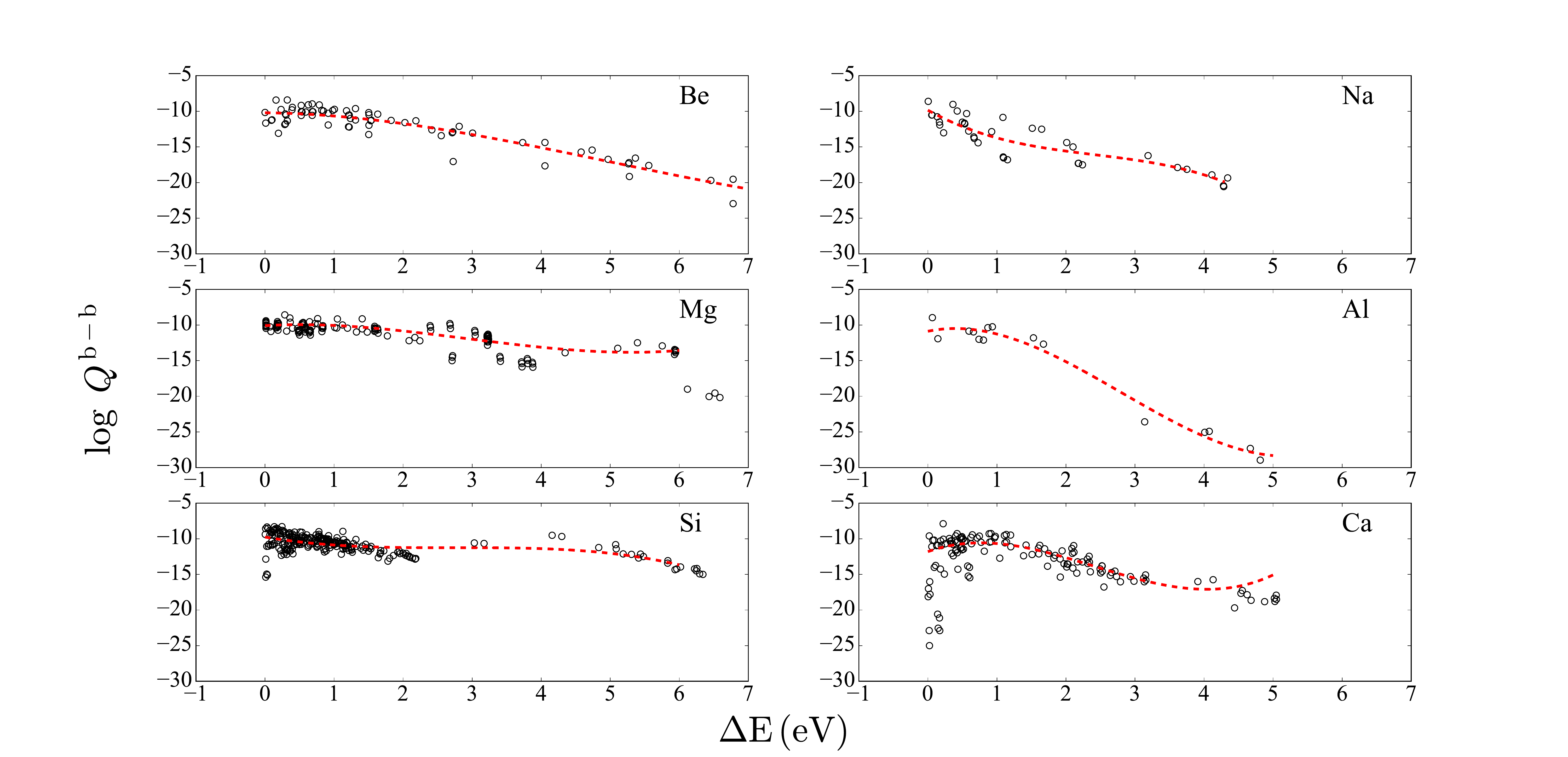}
 \end{center}
 \caption{De-excitation rate coefficients $\log Q^{b-b}$ at $T = 6000$\,K (in logarithmic scale) for \ion{Be}{i}, \ion{Na}{i}, \ion{Mg}{i}, \ion{Al}{i}, \ion{Si}{i} and \ion{Ca}{i} from the quantum calculations of \citet{yakovleva2016}
 \citet{barklem2010}, \citet{belyaev2012} and \citet{guitou2015}, \citet{belyaev2013}, \citet{belyaev2014} and \citet{belyaev2017b}, respectively. Their corresponding polynomial fits (dashed red lines) are also shown.}
 \label{plot_qm_hc_fits_all}
\end{figure*}

\begin{figure*}
\centering
\includegraphics[scale=0.35]{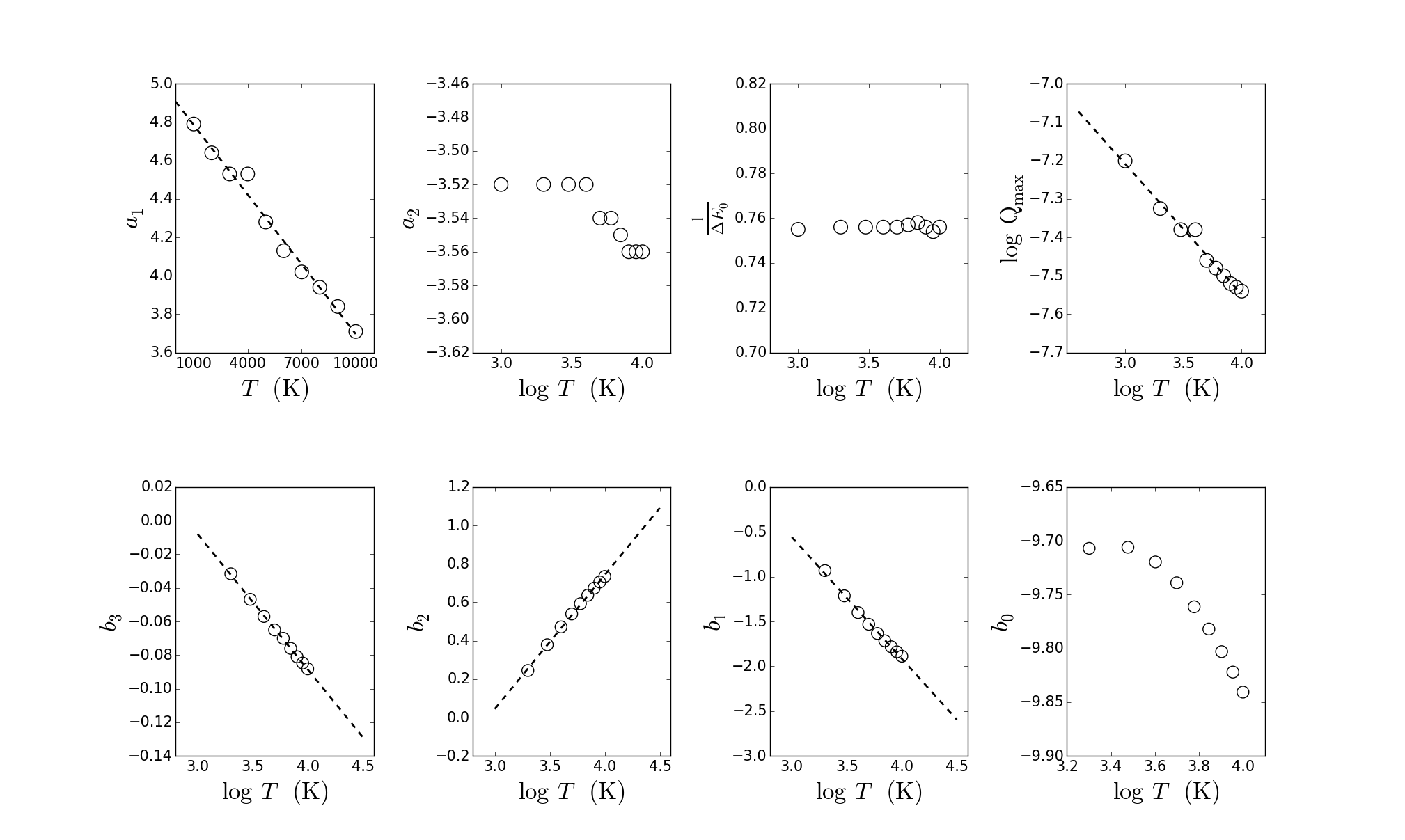}
 \caption{Temperature dependence plots for the CT rate fitting coefficients $a_{1}$, $a_{2}$, $\frac{1}{\Delta E_{0}}$ and $\log Q_{\mathrm{max}}$ (upper panels), and $b-b$ fitting coefficients $b_{1}$, $b_{2}$, $b_{3}$ and $b_{0}$ (lower panels), shown for \sii.}
  \label{param_temp_si}
\end{figure*}

\subsection{De-excitation ($b-b$) rates}\label{de-exc}

The de-excitation (downward $b-b$) rates for \ion{Be}{i}, \ion{Na}{i}, \ion{Mg}{i}, \ion{Al}{i}, \ion{Si}{i} and \ion{Ca}{i} show a larger scatter than CT
rates, with many more transitions; they show a general decrease with increasing transition energy (Figure~\ref{plot_qm_hc_fits_all}).

The rates at large energy differences are in general much smaller than the rates for levels lying close to each other. An exception is Ca for which the collisional  rates between some close lying levels are extremely small. The collisional $b-b$ rates for all six elements can be best fit as a function of $\Delta E$ using a third degree polynomial:

\begin{equation}\label{gen_fit_eqn_hc}
 \log Q^{b-b} = b_{3} \Delta E^{3} + b_{2} \Delta E ^{2} + b_{1} \Delta E + b_{0}
\end{equation}

In addition, $b_{3}$, $b_{2}$ and $b_{1}$ are linearly dependent on $\log T$, while $b_{0}$ does not vary much with temperature, as is demonstrated for Si in the lower panels of 
Figure~\ref{param_temp_si}. This allows us to write the $b-b$ fitting parameters as:
\begin{equation}\label{temp_dep_hc_eqn}
 \begin{cases}
  b_{3} = b_{31} \, \log \frac{T}{10^4} + b_{30}\\
  b_{2} = b_{21} \, \log \frac{T}{10^4} + b_{20}\\  
  b_{1} = b_{11} \, \log \frac{T}{10^4} + b_{10}\\
  b_{0} = b_{0}.\\  
 \end{cases}
\end{equation}

The $b$ parameters for the six elements are listed in Table~\ref{fit_param_all}. The $b_0$ coefficient, that is, the asymptotic value of $\log Q^{b-b}$ for $\Delta\,{\rm E}=0$\,eV, lies between $-9.7$ and $-11.3$, 
with an average $\log Q^{b-b} = -10.40 \pm 0.9$.
The other polynomial fitting coefficients for the $b-b$ rates ($b_{31}$, $b_{30}$, $b_{21}$, $b_{20}$, $b_{11}$ and $b_{10}$) vary more severely from one atom to the next than for the CT rates. This is due to
the large number 
of $b-b$ transitions of differing character. It is also probably affected by the dependency of the rates on a 
double tunneling of the electron between the ionic 
and the valence states, as explained in Section~\ref{ct}. Again, the $b-b$ fitting coefficients' temperature dependence and their effects on the NLTE abundance calculations are tested in Section~\ref{sec:temp_dep}.

\section{Testing the empirical fitting method on Na, Mg, Al and Si atoms} \label{atoms_tests}
The similar behavior of the CT and $b-b$ collisional quantum rates of the six atoms described above prompted us to develop our empirical fitting method to estimate hydrogen collisional rates in NLTE abundance calculations, whenever quantum computations are not available.
To further put our proposed method to the test, we performed NLTE abundance calculations for four of the six atoms described above, Na, Mg, Al and Si. These elements belong to four different groups in the periodic table: alkali metals (Na), alkaline earth metals (Mg), post-transition metals (Al) and metalloids (Si). The other two elements, Be and Ca also belong to the alkaline earth metals and are therefore expected to behave similarly to Mg.

 For that purpose, we built model atoms for Na, Mg, Al and Si, which were used to compute NLTE calculations. We then compared NLTE line profiles and abundance corrections determined for representative lines of each element using hydrogen collisional rates implemented from (i)  published quantum rates and (ii) our empirical fitting method using the fitting coefficients derived for each element from Table~\ref{fit_param_all}. For comparison, we also performed additional calculations for each element using hydrogen collisional rates computed using the classical Drawin equation \citep{drawin1968,drawin1969a} and adopting different fudge factors: (iii) S$_{\mathrm{H}}=0.1$ and (iv) S$_{\mathrm{H}}=1.0$ for \nai, \mgi, \ali\ and \sii, as well as (v) S$_{\mathrm{H}}=0.002$ for \ali\ as recommended by \citet{gehren2004} and \citet{Baumueller1996,Baumueller1997}.
The calculations were tested on a grid of stellar atmospheric parameters corresponding to typical FGK stars.
The model atoms and input model atmospheres used, as well as a detailed explanation of the tests performed to compare the results for each method, are presented below.

\subsection{Model atoms for Na, Mg, Al and Si}\label{model-atoms}
The atoms for Na, Mg, Al and Si used in the tests of the NLTE calculations were homogeneously built by the code \texttt{FORMATO2.0} (Merle et al. in prep) using atomic data adopted from different atomic databases and references as described below.

\begin{table*}
\begin{center}
 \begin{tabular}{r c c c c c c c}
 \hline
 \hline
  Species &  Energy levels & $b-b$ rad. & $b-f$ rad.  & $b-b$ $e^{-}$ col. &  $b-f$ $e^{-}$ col. & $b-b$ H col. & CT H col.\\
  &   & transitions & transitions & transitions & transitions & transitions & transitions\\
 \hline
\nai & 139 & 443 & TOPBASE & SEA62 & SEA62 & BAR10 & BAR10\\
&      NIST & NIST & +KRA68 \\
\hline
\naii & 150 & 141 & TOPBASE & SEA62 & SEA62 & DRW68 & $\cdots$ \\
& NIST & NIST & +KRA68\\
\hline
\mgi & 229 & 1692 & TOPBASE & SEA62+ & SEA62 & BEL12+ & BEL12+\\
     & NIST & VALD3 & +KRA68    & ZAT09 &        & GUI15 & GUI15\\
\hline
\mgii & 82 & 578 & TOPBASE & SEA62 & SEA62 & DRW68 & $\cdots$\\
      & NIST & VALD3 & +KRA68\\
\hline
\ali & 136 & 223 & TOPBASE & SEA62 & SEA62 & BEL13 & BEL13\\
     & NIST+VALD3 & NIST+VALD3 & +KRA68\\
     & +Kurucz database     & +Kurucz database \\
\hline
\alii & 217 & 2320 & TOPBASE & SEA62 & SEA62 & DRW68 & $\cdots$\\
      & NIST & VALD3& +KRA68\\
\hline
\sii & 296 & 9503 & TOPBASE & SEA62 & SEA62 & BEL14 & BEL14\\
     & VALD3 & VALD3 & +KRA68\\
\hline
\sii & 96 & 1019 & TOPBASE & SEA62 & SEA62 & DRW68 & $\cdots$\\
      & VALD3 & VALD3 & +KRA68\\
 \hline 
 \end{tabular}
\end{center}
\caption{Atomic databases and references used for the sources of energy levels and radiative and collisional $b-b$ and $b-f$ transitions used for each species in our model atoms for Na, Mg, Al and Si. Abbreviations correspond to the following references: KRA68 \cite{kramer1968}, SEA62 \cite{seaton1962a}, BAR10 \cite{barklem2010}, DRW68 \cite{drawin1968} and \citet{drawin1969a}, ZAT09 \citet{zatsarinny2009}, BEL12 \citet{belyaev2012}, GUI15 \citet{guitou2015}, BEL13 \citet{belyaev2013}, BEL14 \citet{belyaev2014}.}
\label{tab:model_atoms}
\end{table*}

\subsubsection{Energy levels}\label{si_en_levels}
Fine structure energy levels of the neutral and first ionized species of each element, as well as the mean ground level of the second ionized species, were included in each model atom. Fine-structure energy levels were extracted from the \texttt{NIST}\footnote{https://www.nist.gov/pml/atomic-spectra-database/}, \texttt{VALD3}\footnote{http://vald.astro.uu.se/}, Kurucz atomic line \footnote{http://kurucz.harvard.edu/linelists.html} and The Opacity Project (\texttt{TOPBASE}\footnote{http://cdsweb.u-strasbg.fr/topbase/topbase.html}) databases and the references therein.
The number of fine-structure levels as well as the source databases used for each species of each element are listed in Table~\ref{tab:model_atoms}.

\subsubsection{Radiative transitions}\label{rad_trans_si}
Line data for radiative transitions between
the energy levels were extracted from the \texttt{NIST}, \texttt{VALD3} and Kurucz atomic line databases. The oscillator strengths were adopted from the corresponding references therein.
The Van der-Waals hydrogen collisional broadening coefficients were calculated using the ABO theory \citep{anstee1995,barklem1997,barklem1998} for available transitions, and otherwise using the Uns\"{o}ld approximation \citep{unsold1955}, enhanced by species-dependent fudge factors as recommended by \citet{gustafsson2008}.

All levels in the model atoms were also coupled to the first ionized species' ground levels via bound-free ($b-f$) radiative (photoionization) transitions. The $b-f$ photoionization tables were extracted from the \texttt{TOPBASE} which were calculated using the close-coupling approximation and the R-matrix method. Due to the large number of points in the tables and the sharp resonance peaks in the data, the cross-sections were smoothed and re-sampled before being implemented in the atoms.
For transitions for which no photoionization table exists,
Kramer's hydrogenic approximation \citep{kramer1968} was used to calculate the threshold cross-sections. Details on the number of $b-b$ radiative transitions used for each atom, as well as the database sources used for each atom, are listed in Table~\ref{tab:model_atoms}.

\subsubsection{$e^{-}$ collisional transitions}
All levels in the model atoms are coupled via inelastic collisions with electrons. Quantum calculations for effective collisional strengths $\Upsilon^{e^{-}}$ for electron collisions were implemented for the \mgi\ model atom following \citet{merle2015}. For transitions where no quantum data are available, classical and semi-classical approximations were used. For allowed $b-b$  and ionization by electron collisional transitions, the \citet{seaton1962a} impact parameter approximation was used. For forbidden $b-b$ transitions, \citet{seaton1962b} approximation was used with a collisional strength of $\Upsilon^{e^-}=1$, as recommended for neutral atoms by \citet{allen1973}.

\subsubsection{H collisional transitions}
Inelastic hydrogen collisions were included as follows: different model atoms were built for each element using different recipes for $b-b$ and CT (whenever possible) hydrogen collision rates implemented for the neutral species, using: (i) published quantum rates for $b-b$ and CT collisions (hereafter denoted as the QM models), (ii) our empirical fitting method with fitting coefficients from Table~\ref{fit_param_all} for $b-b$ and CT collisions (hereafter denoted as the EFM model), (iii) Drawin equation for $b-b$ collisions with S$_{\mathrm{H}}=0.1$ and (iv) Drawin equation for $b-b$ collisions with S$_{\mathrm{H}}=1.0$ (denoted as the DRW, S$_{\mathrm{H}}=0.1$ and DRW, S$_{\mathrm{H}}=1.0$ models, respectively). For Al, an additional atom was built implementing H rates using the Drawin equation for $b-b$ collisions with S$_{\mathrm{H}}=0.002$ (denoted as the DRW, S$_{\mathrm{H}}=0.002$). 

For a better comparison of the models, only the transitions for which quantum data are available in the QM models for each element have been implemented in the corresponding EFM models. For the DRW models, we ignored ionization collisions by hydrogen in the DRW models \citep{steenbock1984}, which have been commonly implemented in previous NLTE calculations \citep[e.g.,][]{gehren2004,mashonkina2013}.

In addition to the previous restriction, DRW $b-b$ rates were only included for the collisional transitions with allowed radiative counterparts (i.e., with calculated oscillator strengths).
For the first ionized species (considered dominant in all four elements in FGK-type stars), the Drawin approximation was used.

\subsection{Stellar parameters and model atmospheres}\label{sec:stellar_param}
\begin{table}
\begin{center}
 \begin{tabular}{c| c| c| c}
 \hline
 \hline
 $\log g$ &   $T_{\mathrm{eff}}$ & [Fe/H] & $\xi_{t}$ \\
 (cgs) & (K) & & (\kms)\\
 \hline
 2.5 & 4500 & [$-3.0$,+0.0], $\Delta=1.0$ & 2.0 \\
 3.5 & 4500,5500 & [$-3.0$,+0.0], $\Delta=1.0$ & 1.8\\
 4.5 & 4500,5500,6500 & [$-3.0$,+0.0], $\Delta=1.0$ & 1.5\\
 \end{tabular}
\end{center}
\caption{Grid of stellar parameters for which the NLTE calculations for all atoms were performed.}
\label{tab:stel_param}
\end{table}

To test our empirical fitting method, we perform NLTE calculations for each of the QM, EFM and DRW model atoms of each element for a grid of stellar parameters corresponding to typical FGK-type stars.
The stellar atmospheric parameters used in the grid (effective temperature $T_{\mathrm{eff}}$; surface gravity $\log~g$; metallicity [Fe/H]; and microturbulent velocity, $\xi_{t}$)
are shown in Table~\ref{tab:stel_param}, and were chosen to correspond to three stellar evolutionary stages: giants ($\logg=2.5$), sub-giants ($\logg=3.5$) and dwarfs ($\logg=4.5$). Typical $\xi_{t}$ values were adopted for each \logg value as shown in Table~\ref{tab:stel_param}, such that $\xi_{t}=2.0$ was chosen for the giants, $\xi_{t}=1.8$ for the sub-giants and $\xi_{t}=1.5$ for the dwarfs. Corresponding \Teff values were adopted for each evolutionary stage following the $\Teff-\logg$ diagram from \citet{heiter2015} for the \textit{Gaia} benchmark FGK stars (see their Figure~8). $\Teff=4500$\,K was used for $\logg=2.5,$  $\Teff=4500$\,K and 5500\,K were used for $\logg=3.5,$ and $\Teff=4500$\,K, 5500\,K, and 6500\,K were used for $\logg=4.5$. For each \logg/\,\Teff combination, four metallicity values were adopted from [Fe/H]=$-$3.0 to [Fe/H]=+0.0 in steps of $\Delta=1.0$\,dex. NLTE calculations were thus performed for a total of 24 atmospheric models for each atomic model of each element, giving a total of 96 models for each. One dimensional (1D) LTE MARCS model atmospheres \citep{gustafsson2008} were interpolated to the corresponding atmospheric parameters
using the interpolation routine \texttt{interpol\_marcs.f} written by 
Thomas Masseron\footnote{http://marcs.astro.uu.se/software.php}.

\subsection{NLTE methods}
The NLTE radiative transfer code \texttt{MULTI2.3} \citep{Carlsson1986,carlsson1992} 
was used to compute the level populations and the line EWs. An edited version of \texttt{MULTI2.3} was used\footnote{Corresponding edits were made in the \texttt{mul23\_opacu.f} routine in \texttt{MULTI2.3}.} where CT collisional rates were computed from the input ion-pair production (downward CT) rate coefficients implemented in the model atoms, and the H$^{-}$ number densities determined using the Saha equation \citep{Saha1921}.
Background line opacity files excluding lines of the corresponding elements (e.g., Na background lines were excluded from the calculations for the Na atoms) were included in the
\texttt{MULTI2.3} calculations to account for line blanketing effects. The opacity files were computed for each model metallicity ([Fe/H]) and microturbulent velocity using the MARCS opacity sampling routines.
We computed the corrections for each line, defined as the difference between the NLTE and LTE abundances that corresponds to the same EW, $\Delta\log\varepsilon = \log\varepsilon(\mathrm{NLTE})-\log\varepsilon(\mathrm{LTE})$.

To determine the abundance corrections, we first computed the NLTE and LTE curve-of-growth from five abundance points for each model atom for the 24 atmospheric parameters defined in the grid in Sect.\,\ref{sec:stellar_param}. The abundance points were varied around the central abundance value adopted in each model, in steps of $\Delta=\pm 0.3$\,dex each time. For models where NLTE abundance corrections $>\pm0.6$\,dex were obtained, additional abundance points were added up to $\pm1.2$\,dex from the central abundance to avoid extrapolation.
The abundance corrections between the NLTE
and LTE curves of growth were derived by interpolating
the NLTE curve of growth to the corresponding LTE EW of the median abundance, considered as the reference for each model.
The interpolation method from \citet{steffen1990} was used. Corrections were considered for all lines except for those with $EW<0.5$\,m{\AA}. The NLTE corrections determined for the considered lines of \nai, \mgi, \ali\ and \sii\ at the corresponding \Teff/\logg/[Fe/H] values for all different model atoms described in Section~\ref{model-atoms} are shown in the appendix Tables~A.1. to A.4.

\section{Results}\label{Sec:results}
\begin{figure*}
\begin{center}
\hspace*{-0.5cm}
 \includegraphics[scale=0.22]{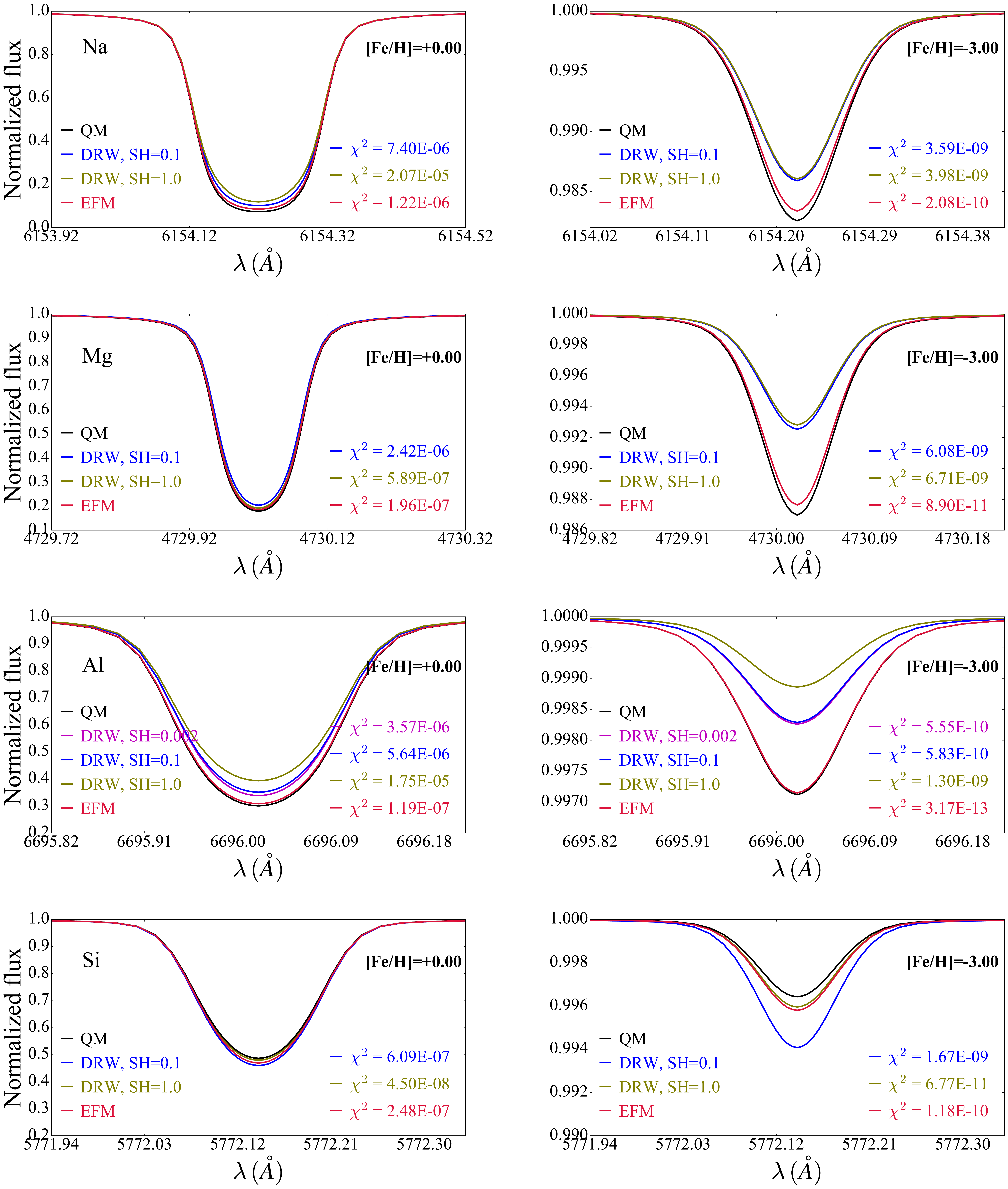}
 \end{center}
 \caption{Line profiles of \nai~6154\,{\AA}, \mgi~4730\,{\AA}, \ali~6696\,{\AA} and \sii~5772\,{\AA} computed using different hydrogen collision models: QM (black lines), EFM (red lines), DRW, S$_{\mathrm{H}}=0.1$ (blue lines), DRW, S$_{\mathrm{H}}=0.1$ (olive lines) and DRW, S$_{\mathrm{H}}=0.002$ for \ali\ (magenta lines). The lines were computed at $\Teff=4500$, $\logg=2.5$ and for two metallicities, $\met=+0.00$ (left panels) and $\met=-3.00$ (right panels). The $\chi^2$ differences obtained between each model relative to the QM are shown on the plots.}
 \label{fig:line-profiles}
\end{figure*}

\begin{figure*}
\begin{center}
\hspace*{-0.5cm}
 \includegraphics[scale=0.23]{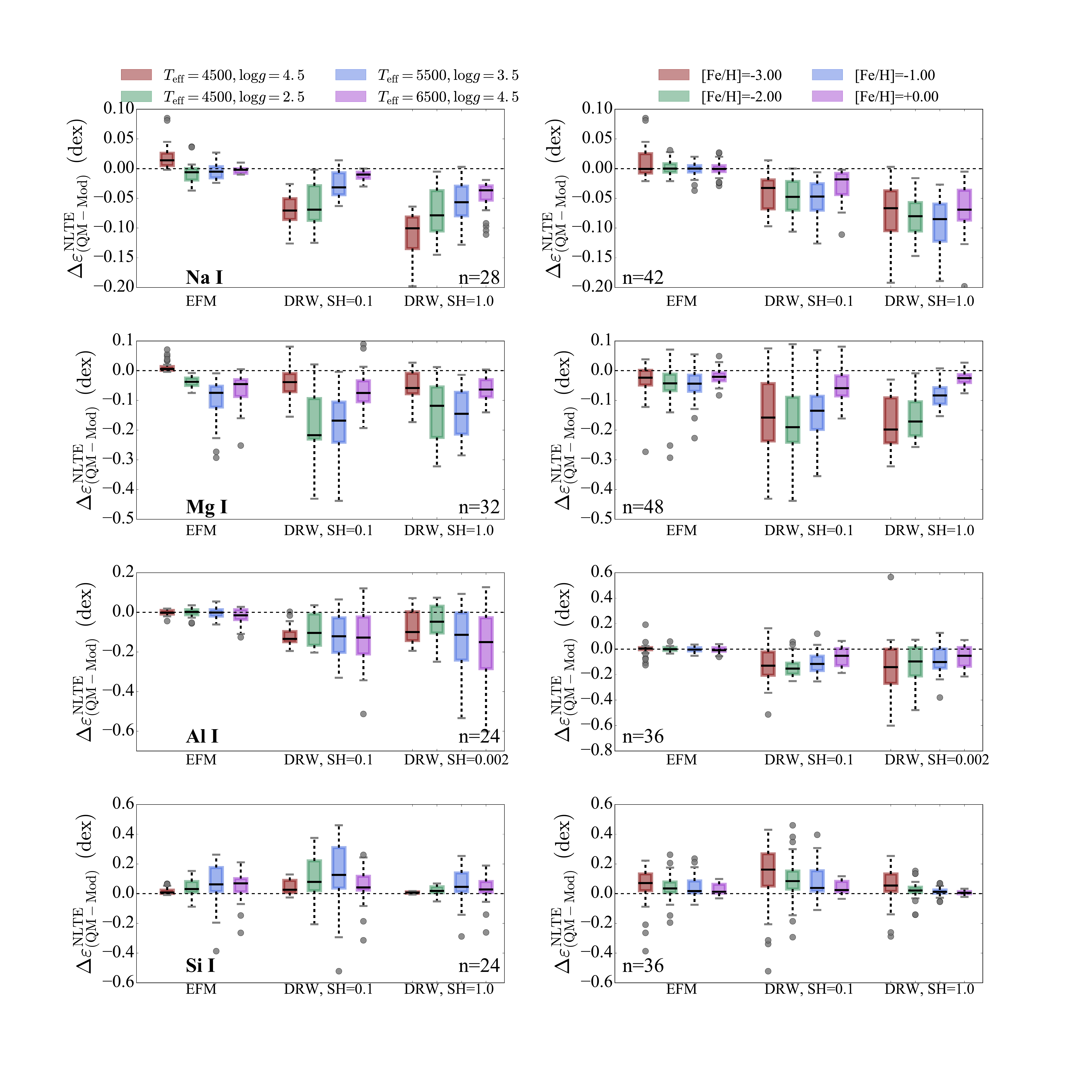}
 \end{center}
 \caption{Box plots showing the overall NLTE abundance correction differences obtained between models computed with the QM rates and those computed with the EFM, DRW, S$_{\mathrm{H}}=0.1$ and DRW, S$_{\mathrm{H}}=1.0$ rates (for \ali, we choose to show results for DRW, S$_{\mathrm{H}}=0.002$ instead of DRW, S$_{\mathrm{H}}=1.0$). Boxes in each panel display the differences obtained from representative lines of \nai, \mgi, \ali\ and \sii\ (shown in Table~A.1.) for the corresponding stellar parameters of cool dwarfs (in red), cool giants (in green), subgiants (in blue) and warm dwarfs (in magenta) in the left-hand panels. Right-hand panels show the same plots for different metallicities. The number of lines, n, used in each box plot is displayed.
 Black solid lines show the median values obtained for each model and each atom, colored boxes represent 50\%, and whisker edges 90\% of the lines. Gray circles show the remaining outlier lines.}
 \label{fig:nlte_corr_boxplot}
\end{figure*}

\begin{figure*}
\begin{center}
\hspace*{-0.7cm}
 \includegraphics[scale=0.22]{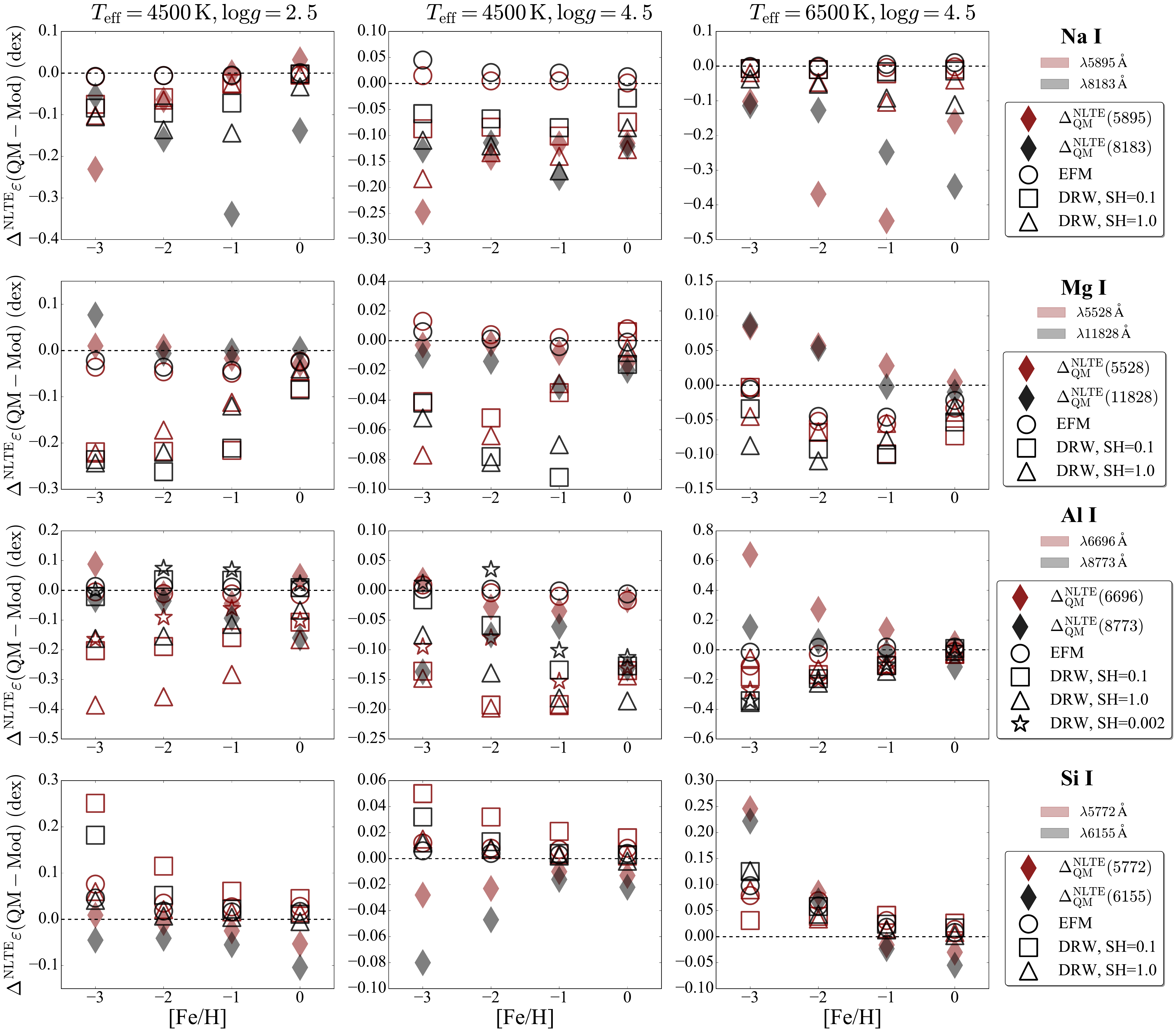}
 \end{center}
 \caption{Abundance correction differences obtained between the QM and the corresponding EFM (empty circles), DRW, S$_{\mathrm{H}}=0.1$ (empty squares) and DRW, S$_{\mathrm{H}}=0.1$ collision models (empty triangles), as well as DRW, S$_{\mathrm{H}}=0.002$ for \ali, (empty stars) for select \nai,  \mgi, \ali\ and \sii\ lines, as a function of $\mbox{[Fe/H]}$. This is shown for the cool giants and dwarfs (left-hand and middle panels), as well as the warm dwarfs (right-hand panels) stellar parameters. The NLTE abundance corrections obtained by the QM models are also shown (filled diamonds), for comparison.}
 \label{fig:nlte_corr_all_1}
\end{figure*}

\begin{figure*}
\begin{center}
\hspace*{-0.7cm}
 \includegraphics[scale=0.22]{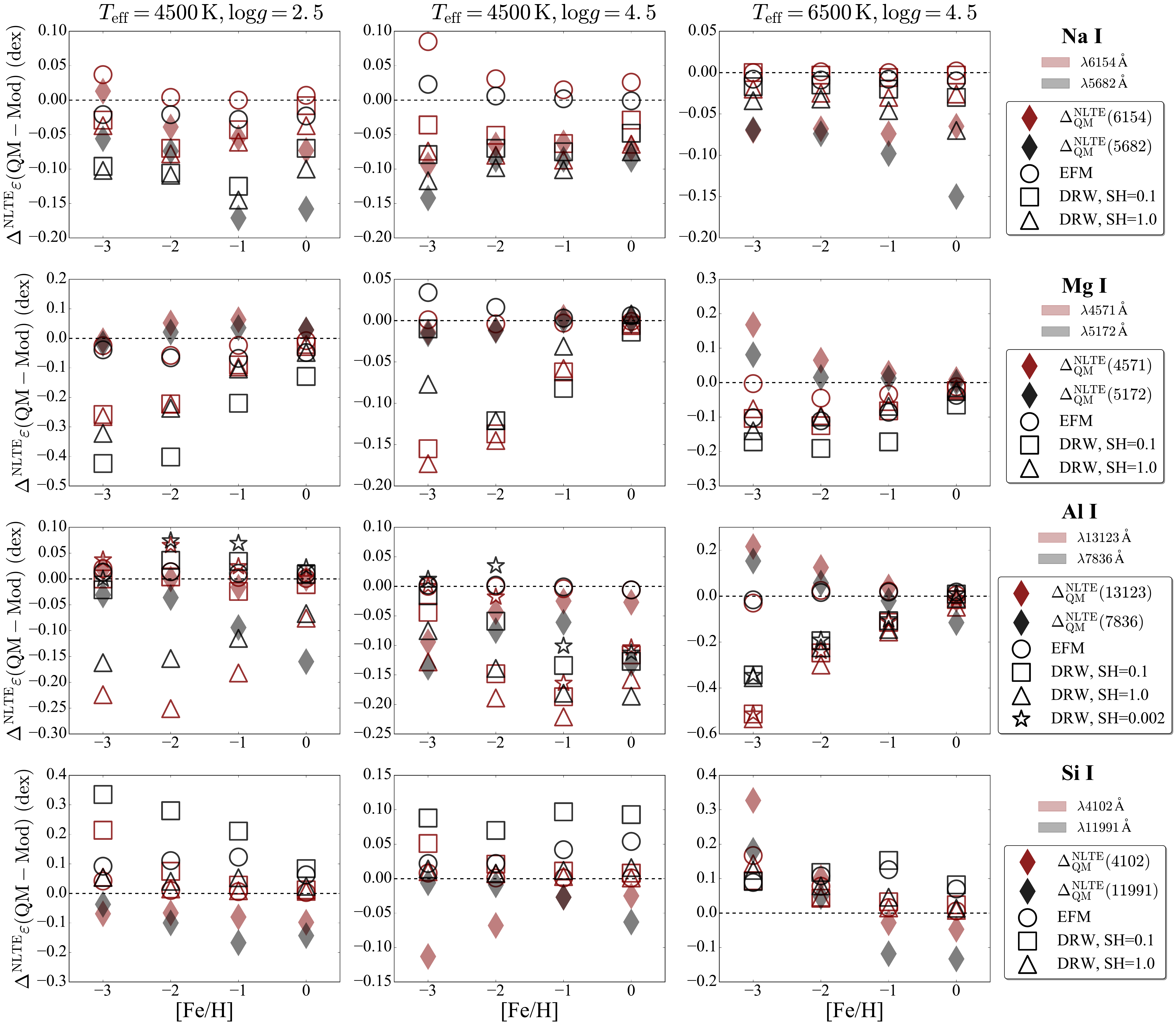}
 \end{center}
 \caption{As in Fig.~\ref{fig:nlte_corr_all_1} but showing different lines for \nai, \mgi, \ali\ and \sii.}
 \label{fig:nlte_corr_all_2}
\end{figure*}

\begin{figure*}
\begin{center}
\hspace*{-0.7cm}
 \includegraphics[scale=0.22]{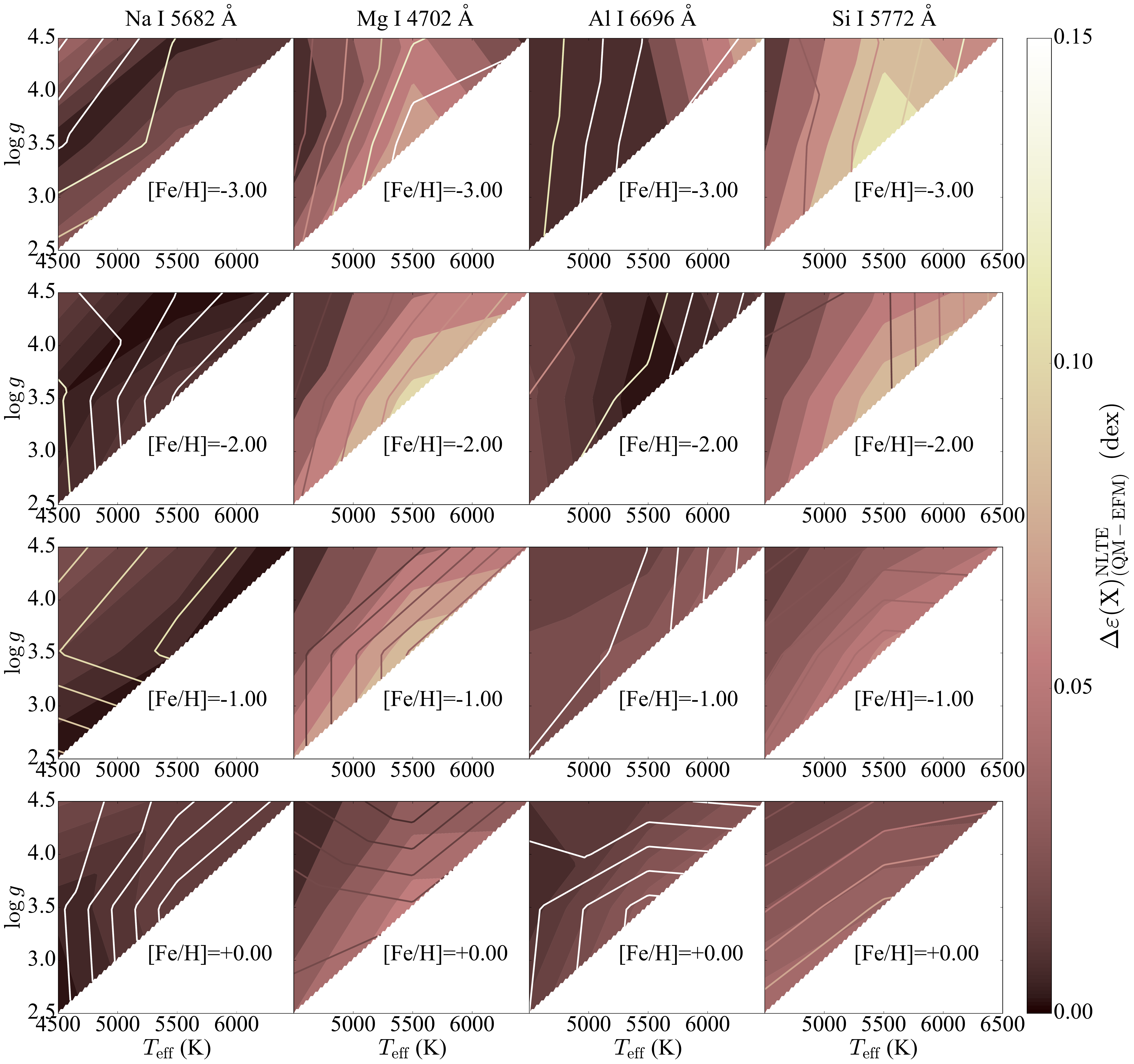}
 \end{center}
 \caption{Abundance correction differences obtained between the QM and our EFM models (in absolute values) for the lines \nai\ 5682\,{\AA},  \mgi\ 4702\,{\AA}, \ali\ 6696\,{\AA} and \sii\ 5772\,{\AA} as a function of \Teff and \logg (represented by the shaded contour plots) and [Fe/H] (upper to lower panels). For comparison, the NLTE abundance corrections (also in absolute values) obtained by the QM models are represented by the iso-contour lines on the same plots using the same color scale.}
 \label{fig:nlte_corr_contour}
\end{figure*}
An important test of the EFM is for its ability to reproduce the line profiles and NLTE abundance corrections computed using the QM, for a range of stellar parameters. We represent below the comparisons of such results obtained with the two collision models for representative lines of \nai, \mgi, \ali\ and \sii. We also compare the line profiles and abundance correction differences obtained between the Drawin models (DRW) to those of the QM.

The aim of this study is to solely compare the NLTE abundance corrections obtained from different implementations of recipes for hydrogen collision rates in the atomic models 
using only the same H collisional transitions as those published for the quantum models. We therefore do not present detailed comparisons to observations or physical interpretation of the obtained NLTE corrections.
We refer the reader to previous detailed NLTE studies for such interpretations such as, but not limited to, \citet{mashonkina2000,andrievsky2007,lind2011} for \nai, \citet{merle2011,mashonkina2013,osorio2015,bergemann2017} for \mgi, \citet{mashonkina2016b,nordlander2017} for \ali\ and \citet{shi2008,bergemann2013,mashonkina2016b} for \sii.

\subsection{Line profile comparisons}
We compare the NLTE line profiles computed using the different collision models. We choose to show in Figure~\ref{fig:line-profiles}, the line profiles computed using the QM, EFM, DRW, S$_{\mathrm{H}}=0.1$ and DRW, S$_{\mathrm{H}}=1.0$ models for the lines at \nai~6154\,{\AA}, \mgi~4730\,{\AA} and \sii~5772\,{\AA}, in addition to the DRW, S$_{\mathrm{H}}=0.002$ model for \ali~6696\,{\AA}. The lines are shown at $\Teff=4500$, $\logg=2.5$ and for both $\met=+0.00$ and $-3.00$, respectively. We compare the lines computed with the different collision models relative to the QM, by calculating the $\chi^2$ difference between each (shown in Figure~\ref{fig:line-profiles}).
The strengths and shapes of the lines computed using the EFM are able to efficiently reproduce those of the QM for the four atomic species. For the solar case ($\met=+0.00$), all collision models compare similarly well relative to the QM. The \nai~6154\,{\AA}, \mgi~4730\,{\AA,} and \ali~6696\,{\AA} lines computed with DRW models have slightly weaker cores than the QM and EFM models. As expected, the lines become much weaker for the extremely metal-poor case ($\met=-3.00$) as compared to the solar one. The line profiles computed using the EFM and QM show very good agreement for the metal-poor stars, while those computed with the DRW models for \nai, \mgi,\ and \ali\ have weaker cores and wings. The \sii~5772\,{\AA} QM line
can be best reproduced with the DRW, S$_{\mathrm{H}}=1.0$ model for both \met\ values, with the EFM models being slightly stronger. We obtained similar results for other lines of \nai, \mgi, \ali\ and \sii\ as well.

\subsection{NLTE abundance correction comparisons}
We then investigate the ability of the EFM to reproduce the results of the QM models by determining the NLTE abundance correction differences between the two collision models for representative lines of \nai\ (7 lines), \mgi\ (8 lines), \ali\ (6 lines) and \sii\ (6 lines) listed in Tables~A.1. to A.4. This was done for the set of stellar parameters defined by our grid (Table~\ref{tab:stel_param}).

\subsubsection{Sodium} 
The overall differences in abundance corrections obtained between the QM and other collision models are shown for \nai\ in the box plots of the upper left panels of Figure~\ref{fig:nlte_corr_boxplot}. Left-hand panels display box plots of the abundance correction differences obtained between the QM and the corresponding collision model including the considered seven \nai\ lines at four different [Fe/H] values (i.e., a total of 28 lines). This is shown for different spectral types defined by the corresponding stellar parameters: cool dwarfs (\Teff=4500, \logg=4.5), cool giants (\Teff=4500, \logg=2.5), subgiants (\Teff=5500, \logg=3.5), and warm dwarfs (\Teff=6500, \logg=4.5). Right-hand panels show similar plots for the differences obtained at different metallicities including all six considered \Teff/\logg\ combinations (i.e., a total of 42 lines).
On average, the \nai\ abundance corrections computed with the EFM models are in agreement with those computed using the QM collisions to within $<0.03$\,dex, at different metallicities and spectral types. Approximately $50\%$ of the lines (defined by the upper and lower limits of the box plots) lie within $\pm0.02$\,dex of the QM. Slightly larger differences, but within $0.04$\,dex, are obtained for the cool dwarfs and the $\met=-3.00$ models. Ninety percent of all lines (defined by the upper and lower limits of the box whiskers) lie within $\pm0.05$\,dex of the reference QM values. Outlier models ($\sim 1\%$ of the lines) show slightly larger differences, up to 0.08\,dex.
These correspond particularly to the \nai\ lines at 6160\,{\AA} and 6154\,{\AA} for the cool dwarf case at $\met=-3.0$. 
The overall differences between the QM and the Drawin models are also displayed. The abundance corrections obtained by the DRW, S$_{\mathrm{H}}=0.1$ models are less discrepant relative to the QM than the S$_{\mathrm{H}}=1.0$. Corrections computed with both DRW models overestimate the QM results by up to $0.10$\,dex for S$_{\mathrm{H}}=0.1$ and $0.15$\, dex for S$_{\mathrm{H}}=1.0$, especially for the cool dwarfs, independent of metallicities.
The effect of collisions in \nai\ NLTE abundance corrections was tested in detail by \citet{lind2011}. They found that the CT hydrogen collisions can be important for cool dwarf models. While our DRW models differ from theirs in terms of ignoring ionization collisions by H, we find similar results to their findings for the cool dwarfs, where the largest differences in abundance corrections between the QM and DRW models are obtained, up to 0.20\,dex.
In general, our results show that our DRW models seem to be able to best reproduce the QM results for the considered \nai\ lines for the warm and solar [Fe/H] stars to within $<0.05$\,dex. This is also in agreement with the results from \citet{lind2011}, who reported negligible hydrogen collision effects on the abundance corrections for solar-type models.

In Figs.~\ref{fig:nlte_corr_all_1} and \ref{fig:nlte_corr_all_2}, we show similar comparisons for four individual \nai\ lines at 5895 and 8183\,{\AA,} and at  5682 and 6154\,{\AA}, respectively,  as a function of [Fe/H], for the cool giants, cool dwarfs, and warm dwarfs.
The corrections determined using the EFM models for the four \nai\ lines are within $<0.05$\,dex
of the QM, for the set of stellar parameters considered. 
We also show for reference, on the same figures, the NLTE abundance corrections obtained by the QM models (denoted by $\mathbf{\Delta^{\mathrm{NLTE}}_{\mathrm{QM}}}$). 
The differences between the QM and EFM models are typically smaller than $\Delta^{\mathrm{NLTE}}_{\mathrm{QM}}$, except for when $\Delta^{\mathrm{NLTE}}_{\mathrm{QM}}\sim0$ where the differences become comparable. 
The corrections computed by the Drawin models, on the other hand, show larger discrepancies ($>0.05$) relative to the QM for the cool star models at lower metallicities ($\met\leq-2.0$).  The discrepancies are smaller for DRW, S$_{\mathrm{H}}=0.1$ and are of the order of $\sim0.1$\,dex relative to the QM. They are, however, in better agreement to within $<0.05$ for the warmer stars and solar-metallicity models. We also show the correction differences between the QM and EFM models (in absolute values) for \nai\ 5682\,{\AA} as a function of \Teff, \logg, and \met\ defined by the contour surface plots of Figure~\ref{fig:nlte_corr_contour}. Values of $\Delta^{\mathrm{NLTE}}_{\mathrm{QM}}$ (also in absolute values) are represented with the line iso-contours on the same figure. Again, the EFM models are able to efficiently reproduce the QM abundance corrections to within $<0.05$\,dex, especially for the higher-gravity models for all four \met\ considered.
The correction differences (shaded contours) between the QM and EFM models are typically smaller than $\Delta^{\mathrm{NLTE}}_{\mathrm{QM}}$ (iso-contours), which can reach up to $\sim 0.15$\,dex for this line at \logg=2.5 for $\mbox{[Fe/H]}=-1.0$ and $\mbox{[Fe/H]}=+0.0$, respectively.

\subsubsection{Magnesium}
Similar to \nai, we compare the abundance correction differences obtained between the EFM and QM models for eight \mgi\ lines at 4571\,{\AA}, 4702\,{\AA}, 4730\,{\AA}, 5172\,{\AA}, 5183\,{\AA}, 5528\,{\AA}, and 5711\,{\AA} and the NIR line at 11828\,{\AA}. Results for 5528\,{\AA} and 11828\,{\AA} are displayed in Figure\,\ref{fig:nlte_corr_all_1}, while those for 4571\,{\AA} and 5172\,{\AA} are shown in Figure\,\ref{fig:nlte_corr_all_2} as a function of stellar parameters.
The \mgi\ abundance corrections determined with the EFM are able to reproduce those of the QM to within $<0.02$\,dex for the cool dwarfs and $<0.05$\,dex for the cool giants. Larger discrepancies exist for the subgiant ($<0.12$\,dex) and warm dwarf cases ($<0.1$\,dex). This is also shown in Figure~\ref{fig:nlte_corr_boxplot} upon considering a total of 32 \mgi\ lines for each spectral type (left panels) and 48 lines for each metallicity (right panels).
The overall differences are smaller for the cool stars. They become slightly larger toward the $\Teff=5500$\,K sub-giants, where the largest discrepancies ($\sim0.3$\,dex) are obtained for \mgi\,5711\,{\AA} at $\met=-3.00$ (shown by the outlier points on Figure~\ref{fig:nlte_corr_boxplot}).
The same behavior is obtained for the four metallicities, with the smallest differences being obtained for the solar values.

For the warm dwarf models, the largest discrepancy of $0.15$\,dex is obtained for the \mgi\ line at 5172\,{\AA}, a component of the \mgi\,b triplets (5167, 5172, 5183). Being very strong at this warm temperature ($\Teff=6500$\,K), this line is, however, rarely used in abundance analyses. The NLTE abundance corrections obtained by the QM models are negligible for the cool stars ($<0.05$\,dex). 
The differences between the QM and our DRW models are more significant for the cool giants and subgiants, with larger discrepancies obtained for S$_{\mathrm{H}}=0.1$. 
In these cases, the abundance correction differences between the DRW and QM models are up to $0.25$\,dex, with some lines even deviating up to 0.40\,dex. We compared our results to those obtained by \citet{mashonkina2013} for the same \mgi\ lines and similar stellar parameters. Our results show that the absolute correction differences between the QM and DRW models are over-estimated as compared to Mashonkina's. For example, they obtain 0.02\,dex and 0.03\,dex absolute differences for the warm dwarfs at $\met=-1.00$ between the QM and DRW, S$_{\mathrm{H}}=0.1$ models for the \mgi\ lines 5172\,{\AA} and 4571\,{\AA}, respectively. We, on the other hand, obtain absolute differences of $0.17$\,dex and $0.08$\,dex, respectively,  for the same lines and stellar parameters. Our values agree better for the 5528\,{\AA} line, with 0.01\,dex difference between our results. Similar discrepancies are obtained for the cool giants. Such differences between both our studies could either be due to (i) neglecting ionization collisions in our DRW models, which leads to stronger absolute NLTE corrections as compared to her results, or (ii) implementing a smaller number of H $b-b$ collisional transitions in our DRW models as compared to her model, where forbidden transitions and those for which no QM rates existed were excluded.
The \mgi\ line abundance correction dependencies on hydrogen collisions were explicitly investigated by \citet{osorio2015}. They found that these dependencies varied from line to line as the level populations in the \mgi\ atom were affected differently by the different H collisional processes. Particularly, the CT rates were found to affect some lines more than others (such as the 5528\,{\AA} line), whereas excitation $b-b$ collisional rates were found to be more important for others (such as for the 8806\,{\AA} line).
Similar to our results for \nai, our results for the DRW models seem to best be able to reproduce the QM abundances for \mgi\ lines at the solar metallicities.

\subsubsection{Aluminum}
For \ali, we consider six representative \ali\ lines, namely, the doublet lines at 6696-6698\,{\AA}, the optical lines at 7836\,{\AA} and 8773\,{\AA}, as well as the NIR lines at 13123\,{\AA} and 16718\,{\AA}. The abundance corrections obtained by the EFM efficiently reproduce those of the QM to within 0.03\,dex. This is shown for \ali\ 6696\,{\AA} and 8773\,{\AA} in Figure~\ref{fig:nlte_corr_all_1}, and for  13123\,{\AA} and 7836\,{\AA} in Figure~\ref{fig:nlte_corr_all_2}. Slightly larger discrepancies of up to $0.05$\,dex are obtained for the warm dwarfs. They are, however, comparatively smaller than the reference QM NLTE abundance corrections which can reach up to 0.6\,dex for the doublet line at 6696\,{\AA} at $\met=-3.0$.

Overall, the corrections derived with the EFM for all our considered \ali\ lines perform well as compared to the QM (within $\pm0.03$\,dex) for different stellar parameters. This is represented by the box plots of Figure~\ref{fig:nlte_corr_boxplot}, when considering 24 \ali\ lines for each spectral type (left panel) and 36 lines for each metallicity (right panel). Unlike the DRW models, the EFM is better able to reproduce the QM abundances of \ali\ lines at the lowest metallicities. 

The Drawin models can overestimate the abundances relative to the QM up to $0.4$\,dex for S$_{\mathrm{H}}=1.0$ for the cool giants and $0.2$\,dex for the dwarfs. \citet{nordlander2017} also compared their abundance corrections computed using the QM rates to those computed using the DRW, S$_{\mathrm{H}}=0.1$ models by \citet{andrievsky2008}. They found correction differences up to 0.3\,dex between the two studies for the dwarf stars. We compare our NLTE abundance corrections obtained with the QM models to those obtained by \citet{nordlander2017} and find agreement in our results to within 0.1\,dex for similar lines and stellar parameters, even though the authors implemented slightly different atomic data including more transitions for both electron and hydrogen collisions. Comparing our corrections obtained by the QM models to those computed by \citet{mashonkina2016b}, we see slightly larger discrepancies up to 0.2\,dex. These discrepancies could, again, be due to the differences in our atomic models. These latter authors used the classical \citet{vanregemorter1962} approximation for their electron $b-b$ collision rates, whereas the semi-empirical \citet{seaton1962a} collisional recipe was used in ours.
In addition to S$_{\mathrm{H}}=0.1$ and S$_{\mathrm{H}}=1.0$, we also consider S$_{\mathrm{H}}=0.002$ for \ali\ as recommended by previous studies \citep{Baumueller1996,Baumueller1997}. The latter performs slightly better than the larger scaling factors at reproducing the QM corrections for some lines for the cool giants and dwarfs. That is not the case, though, for the warm dwarfs where S$_{\mathrm{H}}=0.002$ produces larger discrepancies up to $0.4$\,dex at $\met=-3.0$. These differences between our QM and DRW models could be due to the contributing effects of CT collisions in \ali, in addition to ignoring the hydrogen ionization collisions in our \ali\ DRW models as compared to previous NLTE studies, which could be driving the larger discrepancies relative to the QM. The discrepancies between the Drawin and QM models are also smaller at the solar metallicities, similar to what was obtained for \nai\ and \mgi.

\subsubsection{Silicon}
For \sii, we consider six commonly used lines in spectroscopic studies, namely, 4102\,{\AA}, 5772\,{\AA}, 6155\,{\AA} and 8752\,{\AA} as well as two NIR lines at 11984\,{\AA} and 11991\,{\AA}.
The EFM corrections slightly underestimate those computed with the QM, by $0.05$\,dex for the cool giants, $0.02$\,dex for the cool dwarfs, and up to $0.1$\,dex for the warm dwarfs at $\met\leq-2.0$. These discrepancies between the EFM and the QM are furthermore highlighted for the \sii\ line at 5772\,{\AA} in Figure~\ref{fig:nlte_corr_contour}, as function of \Teff, \logg, and \met. Indeed, the differences are pronounced at $\met=-3.0$, especially toward the subgiant gravities. 
The NLTE abundance corrections obtained by the QM models, however, are small and are typically $-0.05$\,dex for the cool giants and $-0.1$\,dex for the cool dwarfs.

The abundances determined with the DRW, S$_{\mathrm{H}}=1.0$ are similar to those obtained by the EFM and QM for different stellar parameters. The independence of the \sii\ NLTE corrections on the different H collisional models has also been shown in the previous studies of \citet{bergemann2013} and \citet{mashonkina2016b}.

\section{Removing the temperature dependence from the EFM}\label{sec:temp_dep}
\begin{figure}
\begin{center}
\hspace*{-0.7cm}
 \includegraphics[scale=0.22]{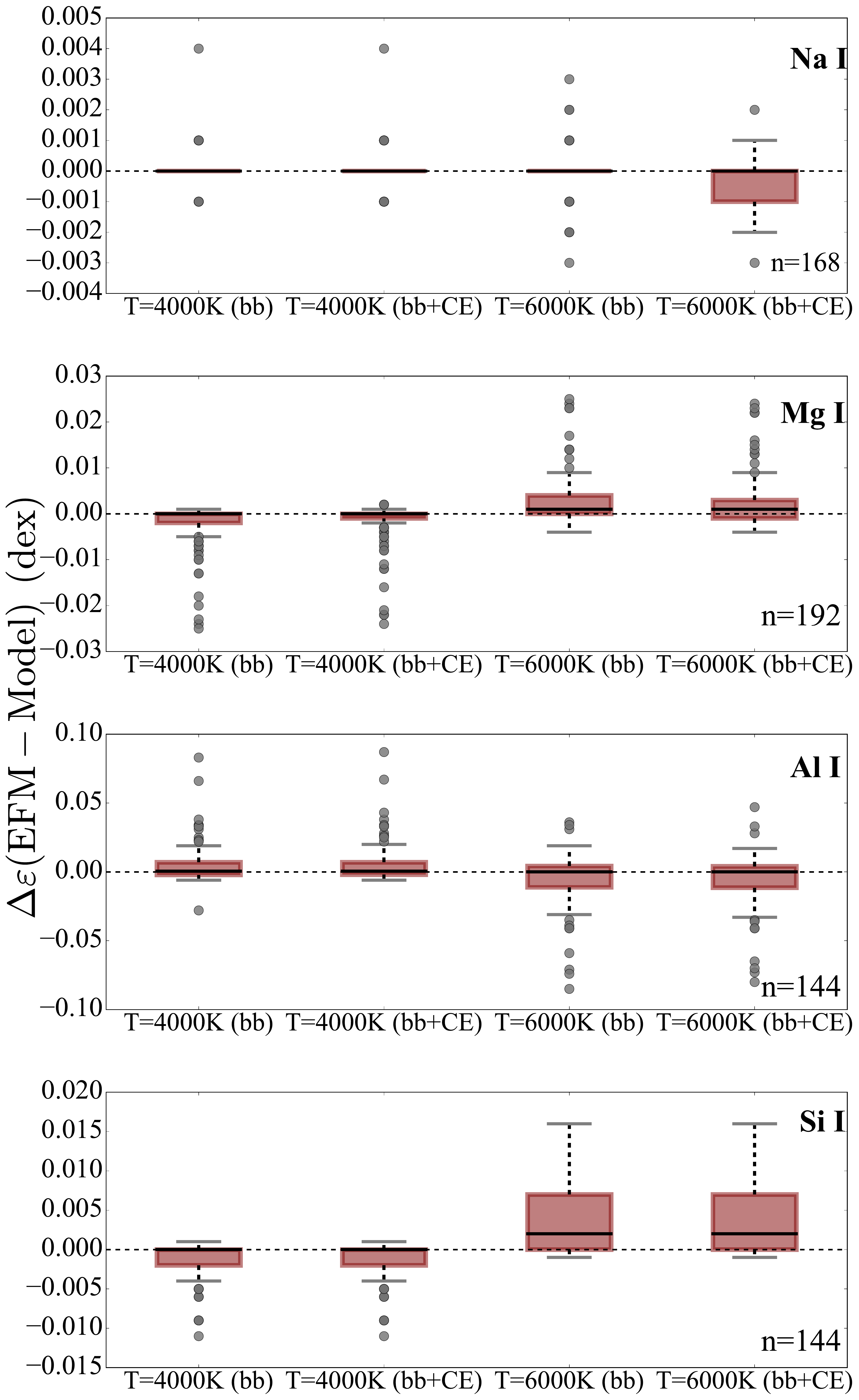}
 \end{center}
 \caption{Box plots representing the NLTE abundance correction differences obtained between models computed with $T$-dependent EFM rates and those computed with $T$-independent EFM rates for either the $bb$ or both $bb$+CE collisions at $T=4000$\,K and $T=6000$\,K, respectively. The different panels show the results obtained at each model considered for each of the \nai, \mgi, \ali\ and \sii\ representative lines (Tables~A.1 to A.4) at all considered stellar parameters (Table~\ref{tab:stel_param}). The number of lines, n, used by each model is also displayed on the plots. The solid black lines in the boxes show the median differences obtained at each model for each atom. Colored boxes represent 50\%, and whisker edges (dotted black lines) 90\% of the lines. Gray circles show the remaining outlier lines}
 \label{fig:tconst_comp}
\end{figure}

Finally, we test for the possible effects of removing the temperature dependencies of the EFM hydrogen collision rates from Eqs. \ref{temp_dep_ce_eqn} and \ref{temp_dep_hc_eqn}, respectively. This is motivated by the fact that the fitting parameters depend weakly on temperature, as discussed in Sects.\,\ref{ct} and \ref{de-exc}. Another motive is the possibility to decrease the number of fitting coefficients of the EFM from six for the CT and seven for the $b-b$ rates, to four fitting parameters for each.

Applying the EFM to derive hydrogen collisional rates for atoms with no published quantum rates would require deriving these fitting parameters for the required atom by $\chi^{2}$ minimization fitting to spectra or measured EWs of benchmark or reference stars for a range of stellar parameters. This would require computing large grids of models for different combinations of fitting coefficients for each of the CT and $b-b$ recipes in Eqs.\,\ref{gen_fit_eqn_ce} and \ref{gen_fit_eqn_hc}. Decreasing the number of fitting parameters would typically simplify the process of deriving these coefficients. 
We therefore perform additional tests by computing NLTE abundances for the lines of each element using collision rates via the EFM with (i) constant temperature $T=4000$\,K for $b-b$ rates, (ii) constant temperature $T=4000$\,K for both $b-b$ and CT rates, (iii) constant temperature $T=6000$\,K for $b-b$ rates, and finally (iv) constant temperature $T=6000$\,K for both $b-b$ and CT rates. This was done again for the set of stellar parameters defined by our grid in Table~\ref{tab:stel_param}. 

We compared the NLTE abundance corrections computed with each of these four models for each element to those calculated with $T$-dependent EFM coefficients. The differences obtained from all the lines defined in Section~\ref{atoms_tests} at all stellar parameters are represented by the box plots for each element in Fig.~\ref{fig:tconst_comp}.
The discrepancies between different $T$=constant models and the EFM are negligible for \nai\ lines, with  differences of approximately zero (with a few outlier lines lying at $<0.004$\,dex). Using constant $T=6000$\,K collision rates for both $b-b$ and CT collisions results in slightly larger but equally negligible differences up to $-0.002$\,dex.
For \mgi, the differences of most lines are within $<0.005$\,dex for $T=4000$\,K and $<0.01$ for $T=6000\,K$, with a few outliers at $<0.02$\,dex. For \ali, the constant $T$ collisions lead to slightly larger corrections than the $T$-dependent EFM models to a maximum value of $0.02$\,dex, especially at $T=6000$\,K. A few outliers even deviate by up to $-0.1$\,dex from the EFM. Finally, for \sii\ lines, the corrections determined with the $T$-independent collisions differ slightly from the $T$-dependent ones within $-0.005$\,dex at $T=4000$\,K and $+0.01$\,dex at $T=6000$\,K.

Overall, these results show that the NLTE calculations do not vary strongly by removing the $T$ dependence from H collisional rates, except for a few outliers. This is especially the case upon using a constant $T=4000$\,K for both CT and $b-b$ rates. The temperature dependencies can therefore be ignored in Eqs.\,\ref{ct} and \ref{de-exc}. This allows the number of fitting coefficients of our proposed EFM model to be decreased to four for  CT and for $b-b$ rates, which can potentially be used to derive rate coefficients for elements for which no quantum rates yet exist. This is yet to be tested in future work.


\section{Conclusions}\label{conclusion}

We introduce an empirical fitting method (EFM) to estimate hydrogen collision rates to be used in NLTE abundance calculations. The method is based on fitting the published quantum rates of the neutral species for elements such as Be, Na, Mg, Al, Si and Ca with simple polynomial functions, whose rates show a very similar behavior as a function of transition energies. We provide two general fitting recipes, one for charge transfer rates, and one for de-excitation rates, that can be used for all six elements. Sets of six and seven coefficients for  $b-b$ and CT collision processes, respectively,  define the fits. 
The coefficients were found to vary slightly from one element to the next, but the variations remain within a range that suggest generality of the recipe. This implies that the recipes can potentially be used for elements with unknown hydrogen collision rates. The temperature dependence of the rates was taken into account, however, abundances were found to depend weakly on collision temperatures, which in principle can be removed from the fits. This decreases the number of fitting coefficients to four for each collisional process.

The fitting recipes were tested for their ability to reproduce the NLTE line profiles and abundance corrections of models incorporating quantum collisional rates (QM). These were determined for typical lines of four atoms for Na, Mg, Al and Si, over a grid covering a range of stellar parameters typical for FGK stars. 
For \nai\ lines, the NLTE corrections determined using the EFM are able to reproduce those computed with the QM rates to within $\pm0.03$\,dex for the warm dwarfs and subgiants, and to within 0.02\,dex for the cool giants and dwarfs. 
For \mgi\ and \sii\ lines, while abundances for most lines lie within $0.05$\,dex of the QM abundances, larger differences were obtained for some cases up to 0.15\,dex for \mgi, and 0.2\,dex for \sii. This was especially the case for the subgiants and warm dwarfs. Smaller differences of $<0.05$\,dex were obtained for the cool dwarfs and giants. 
The EFM performs well in reproducing the QM abundance corrections for the \ali\ lines. The majority of considered \ali\ lines lie within $0.03$\,dex of the reference QM models for different stellar atmospheric models. This is especially important for the metal-poor stars, where \ali\ NLTE abundance corrections can reach up to 0.6\,dex at $\met=-3.0$.

We also compared the abundance corrections obtained with the QM and EFM models to those calculated with the Drawin equation (DRW), incorporating the same excitation transitions as the QM, while ignoring ionization transitions by hydrogen.
The absolute correction differences obtained between the QM and DRW models are larger in magnitude than those obtained between the QM and EFM models by up to  0.4\,dex. This is especially the case for \nai, \mgi\ and \ali\ at lower metallicities. 
The discrepancies between the QM and DRW models at the optimal S$_{\mathrm{H}}$ values for each species (0.1 for \nai, 1.0 for \mgi, 0.002 and 0.1 for \ali\ and 1.0 for \sii) are typically smallest for the solar metallicities. 
For the \sii\ lines, the corrections determined with the EFM are comparable to those computed with the DRW models for S$_{\mathrm{H}}=1.0$. This agrees with previous studies for \sii\ \citep{bergemann2013,mashonkina2016b} which showed that NLTE abundance corrections depend weakly on hydrogen collisions. Caution is advised, however, when comparing the abundance corrections derived by our DRW models to those determined in previous studies, as less H excitation and no ionization collisional transitions were incorporated in our models. This could potentially be driving the larger discrepancies obtained relative to the QM models, as compared to the literature.

Our proposed empirical fitting method, tested on four different neutral elements belonging to different groups from the  period table, is able to reproduce the line profiles and NLTE abundance corrections of the same atoms incorporating ab-initio data, to within the abundance precisions described above for the different elements. It generally performs best for the cool and warm dwarfs, with slightly larger discrepancies obtained for the cool giants and subgiants. Future tests should exploit its ability to estimate the rates for energy levels that have not been included in the ab-initio calculations, and moreover, even potentially apply it to derive rates for other elements for which no quantum rates have yet been published.

\begin{acknowledgements} 
We thank the anonymous referee for insightful comments and suggestions that have helped significantly  to improve our paper.  
We thank Paul Barklem for valuable discussions on inelastic hydrogen collisions throughout the past years. Matthias Steffen is also thanked for making available his interpolation package. R.E. acknowledges support from a JINA-CEE fellowship, funded in part by the National Science Foundation under Grant No. PHY-1430152. T.M. and M.V.d.S.are supported by a grant from the Fondation ULB.
This research was partially financed by the CEDRE PHC program, and the CNRS Programme National de Physique Stellaire. This work has made use of the VALD database, operated at Uppsala University, the Institute of Astronomy RAS in Moscow, and the University of Vienna and the Atomic Spectra Database hosted by the National Institute of Standards and Technology (NIST). 

\end{acknowledgements}


\begin{appendix}
\section{Line data - NLTE corrections}
The tables in this Appendix, A.1. - A.4. listing the NLTE corrections obtained for selected lines of \nai, \mgi, \ali\ and \sii\ are only available in electronic form at the CDS via anonymous ftp to cdsarc.u-strasbg.fr (130.79.128.5) or via http://cdsweb.u-strasbg.fr/cgi-bin/qcat?J/A+A/. The corrections are computed with four different atoms for each element of different implementations of hydrogen collisions using (i) the published quantum rates (QM), (ii) our proposed fitting method (EFM), (iii) the Drawin equation with S$_{\mathrm{H}}=0.1$ (DRW, S$_{\mathrm{H}}=0.1$) (iv) the Drawin equation with S$_{\mathrm{H}}=1.0$ (DRW, S$_{\mathrm{H}}=1.0$), as well as (v) the Drawin equation with S$_{\mathrm{H}}=0.002$ (DRW, S$_{\mathrm{H}}=0.002$) for \ali, for a range of stellar parameters defined in Table~\ref{tab:stel_param}.

\end{appendix}

\end{document}